\documentstyle[preprint,aps]{revtex}
\tighten
\draft
\widetext
\preprint{YITP-00-5}
\bigskip
\bigskip
\begin{document}
\title{Localized (Super)Gravity and Cosmological Constant}
\medskip
\author{Zurab Kakushadze\footnote{E-mail: 
zurab@insti.physics.sunysb.edu}}
\bigskip
\address{C.N. Yang Institute for Theoretical Physics\\ 
State University of New York, Stony Brook, NY 11794}

\date{May 24, 2000}
\bigskip
\medskip
\maketitle

\begin{abstract}
{}We consider localization of gravity in 
domain wall solutions of 
Einstein's gravity coupled to a scalar field with a generic
potential. We discuss conditions on the scalar
potential such that domain wall solutions are non-singular.
Such solutions even exist for appropriate
potentials which have no minima at all and are unbounded below. 
Domain walls of this type have infinite tension, while usual kink type of
solutions interpolating between two AdS minima have finite tension. In the
latter case the cosmological constant on the domain wall is necessarily
vanishing, while in the former case it can be zero or negative. 
Positive cosmological constant is allowed for singular domain walls. We
discuss non-trivial conditions for physically allowed singularities arising
from the requirement that truncating the space at the singularities be
consistent. Non-singular domain walls with infinite tension 
might {\em a priori} avoid recent ``no-go'' theorems indicating 
impossibility of supersymmetric
embedding of kink type of domain walls in gauged supergravity. We argue that
(non-singular) domain walls are stable even if they have infinite tension. This
is essentially due to the fact that localization of gravity in smooth domain
walls is a Higgs mechanism corresponding to a spontaneous breakdown of
translational invariance. As to
discontinuous domain walls arising in the presence of $\delta$-function 
``brane''
sources, they explicitly break translational invariance. Such solutions cannot
therefore be thought of as limits of smooth domain walls. 
We point out that if the
scalar potential has no minima and approaches finite negative values at
infinity, then higher derivative terms 
are under control, and do not affect the 
cosmological constant which is vanishing for such backgrounds. 
Nonetheless, we also point out that 
higher curvature terms generically delocalize gravity, so that the desired
lower dimensional Newton's law is no longer reproduced. 
\end{abstract}

\section{Introduction}

{}In the Brane World scenario the Standard Model gauge and matter fields
are assumed to be localized on $p\leq 9$ spatial dimensional 
branes (or an intersection thereof), while gravity lives in a larger
(10 or 11) dimensional bulk of space-time 
\cite{early,BK,polchi,witt,lyk,shif,TeV,dienes,3gen,anto,ST,BW}. The volume
of dimensions transverse to the branes is then assumed to be finite. This is
automatically achieved if the transverse dimensions are compact. However,
as was pointed out in \cite{RS}, if one considers certain 
warped \cite{Visser} compactifications, then the volume of the 
transverse dimensions
can be finite even if the latter are non-compact. Another way of phrasing this
result is that gravity is localized on a lower dimensional subspace (that is, 
the brane)\footnote{The idea to use warped compactifications to achieve
localization of gravity was originally presented in \cite{Gog}. In \cite{RS}
a concrete realization of this idea was given in the context of one extra
dimension. In \cite{many} this was generalized to cases with more extra
dimensions. In \cite{ver} possible string theory 
embeddings of such a brane world scenario were discussed. For a 
general discussion
of localization of various fields in such backgrounds, see \cite{giga}.}.

{}One motivation for considering such unconventional compactifications is 
the moduli problem. In particular, the extra dimensions in such scenarios are
non-compact while their volume is finite and fixed in terms of other 
parameters in the theory such as those in the scalar potential. That is, the
scalars descending from the components of the higher dimensional metric 
corresponding to the extra dimensions are actually massive, and their 
expectation values are fixed.

{}Also, scenarios with localized gravity could possibly have implications for
the cosmological constant problem. In particular, in this context together with
the localized graviton zero mode one has a continuum of massive bulk graviton  
modes. One can then ask whether this can somehow help avoid the usual 
four-dimensional field theory arguments \cite{weinberg} why having 
vanishing (or small) cosmological constant in the absence of supersymmetry 
generically seems to require gross fine-tuning. Thus, on one hand, 
{\em a priori} it is not
completely obvious whether the four-dimensional effective field theory 
arguments
do always apply in this context. On the other hand, the fact that the 
volume of extra dimensions is finite seems to suggest that they should, at 
least at low enough energies. One of the purposes of this paper is to get 
additional insights into these issues.

{}In this paper we consider localization of gravity arising in 
$(D-1)$-dimensional domain wall solutions in the system of $D$-dimensional
Einstein gravity coupled to a single real scalar field with a generic
scalar potential. In particular, we discuss conditions on the scalar
potential such that the corresponding domain wall solutions are non-singular
(in the sense that singularities do not arise at finite values of the 
coordinate transverse to the domain wall). The usual kink type of solutions
are non-singular as they interpolate between two adjacent local AdS minima
of the scalar potential. Such solutions always have vanishing
$(D-1)$-dimensional cosmological constant. On the other hand, we point out that
there exist other non-singular solutions (subject to the aforementioned 
non-singularity conditions on the scalar potential) which do {\em not}
interpolate between AdS minima. In fact, such solutions exist even for
potentials which have no minima at all and are unbounded below. Domain walls of
this type have infinite tension. The $(D-1)$-dimensional cosmological constant
in such solutions can be vanishing or negative. However, positive cosmological
constant is allowed for singular solutions. In fact, we discuss non-trivial
conditions for physically allowed singularities, which arise from the 
requirement that truncating the space at the singularities be consistent.
In particular, as we point out in the following, these conditions are not
satisfied in some of the recently discussed ``self-tuning'' solutions.

{}One possible implication of the existence of non-singular solutions with
infinite tension is that if the corresponding potentials can be obtained
from, say, $D=5$ ${\cal N}=2$ gauged supergravity, then one would obtain a 
supersymmetric domain wall 
with localized supergravity. In particular, {\em a priori} it is
unclear whether such potentials avoid 
recent ``no-go'' theorems, which seem to imply that 
potentials with more then one AdS minima cannot be obtained in this context
thus ruling out supersymmetric kink type of solutions. 
If not, then it would be important to show
that the aforementioned ``no-go'' theorems also extend to infinite tension
domain walls.

{}In this paper we also study the issue of stability of domain walls which
localize gravity. In particular, {\em a priori} it is not obvious why, say,
infinite tension domain walls, arising in theories with potentials which
have no minima at all and are unbounded below, should be stable. 
Indeed, the $D$-dimensional
theory in such cases appears to be sick, and could even have tachyons. 
Nonetheless, we argue that (non-singular) $(D-1)$-dimensional solutions
are (classically) stable even for non-vanishing $(D-1)$-dimensional 
cosmological constant (in particular, in the latter case the corresponding 
domain walls are not BPS saturated). The basic reason for stability is that
in the case of smooth domain walls (that is, those without any {\em ad hoc}
$\delta$-function ``brane'' sources which break $D$-dimensional diffeomorphism
invariance explicitly), localization of gravity is a Higgs mechanism for the
graviton field in the process of which the scalar field is eaten 
(or, more precisely, its corresponding modes are) by the would-be massless
$(D-1)$-dimensional graviphoton which acquires non-zero mass. In particular, 
we point out the
importance of the full $D$-dimensional diffeomorphism invariance for the
self-consistency of this Higgs mechanism.

{}In the light of the above discussion it is appropriate to mention that
discontinuous domain walls arising in the presence of 
$\delta$-function ``brane'' sources {\em cannot} be though of as limits of
smooth solutions. One way to understand this is to note that such ``brane''
sources break translational invariance explicitly, while in the case of smooth
domain walls this breakdown is spontaneous. In particular, there is an explicit
discontinuity between these two different setups unless the $(D-1)$-dimensional
Planck scale is vanishing.

{}At the end of the paper we discuss some aspects of the cosmological constant
problem in the context of a brane world realized via a domain wall which
localizes gravity. In particular, we point out that higher derivative such
as higher curvature terms do not seem to be under control unless we consider
the following type of domain walls. Consider a potential with no local minima
such that it approaches finite negative values for infinite values of the 
scalar field. In this case we argue that the higher derivative terms are
under control (as long as the AdS curvature at infinity is small enough 
compared with the cut-off scale for the higher derivative terms). Then 
provided that quantum corrections do not modify the aforementioned behavior
of the scalar potential (that is, if the scalar potential still approaches
constant negative values at infinity), one might hope to solve the 
cosmological constant problem in this context as the $(D-1)$-dimensional 
cosmological constant in such solutions is ultimately zero
(albeit it is not completely
clear how to control the local corrections which could modify the behavior
of the scalar potential at infinity).

{}Even though the above setup might appear promising in the
context of the cosmological constant problem, we point out certain difficulties
with such a scenario, which, in fact, seem to be generic to theories with
finite volume non-compact extra dimensions. In particular, even though 
higher curvature terms appear to be under control as far as preserving the 
desirable properties of the
corresponding warped background is concerned, they generically {\em delocalize}
gravity. That is, once we include such terms, generically there is no longer a
normalizable
graviton zero mode, and the usual $(D-1)$-dimensional Newton's law is no
longer reproduced. We also point out some possibilities for avoiding these
problems, which will be discussed in more detail elsewhere.  

\section{Setup}

{}In this section we discuss the setup within which we will discuss
solutions with localized gravity. Thus, consider a single real 
scalar field $\phi$
coupled to gravity with the following action\footnote{Here we focus on
the case with one scalar field for the sake of simplicity. In particular,
in this case we can absorb a (non-singular) metric $Z(\phi)$ in
the $(\nabla\phi)^2$ term by a non-linear field redefinition. This cannot
generically be done in the case of multiple scalar fields $\phi^i$, where
one must therefore also consider the metric $Z_{ij}(\phi)$.}:
\begin{equation}\label{action}
 S=M_P^{D-2}\int d^D x \sqrt{-G}\left[R-{4\over{D-2}}(\nabla\phi)^2 -V(\phi)
 \right]~,
\end{equation} 
where $M_P$ is the $D$-dimensional (reduced) Planck mass. The equations of 
motion read:
\begin{eqnarray}
 && {8\over{D-2}}\nabla^2\phi- V_\phi=0~,\\
 &&R_{MN}-{1\over 2}G_{MN} R={4\over {D-2}}\left[\nabla_M\phi\nabla_N\phi
 -{1\over 2}G_{MN}(\nabla \phi)^2\right]-{1\over 2}G_{MN} V~.
\end{eqnarray}
The subscript $\phi$ in $V_\phi$ denotes derivative w.r.t. $\phi$.

{}In the following we will be interested in solutions to the above equations
of motion with the warped \cite{Visser} metric of the following form:
\begin{equation}\label{warped}
 ds_D^2=\exp(2A)ds_{D-1}^2+dy^2~,
\end{equation}
where $y\equiv x^D$, the warp factor $A$, which is a function of $y$,
is independent of the coordinates
$x^\mu$, $\mu=1,\dots,D-1$, and the $(D-1)$-dimensional interval is given
by
\begin{equation}
 ds_{D-1}^2={\widetilde g}_{\mu\nu}dx^\mu dx^\nu~,
\end{equation}  
with the $(D-1)$-dimensional metric 
${\widetilde g}_{\mu\nu}$ independent of $y$. 

{}With the above ans{\"a}tz, we have:
\begin{eqnarray}
 &&R_{\mu\nu}={\widetilde R}_{\mu\nu}-\exp(2A)\left[A^{\prime\prime}+
 (D-1)(A^\prime)^2\right] {\widetilde g}_{\mu\nu}~,\\
 &&R_{DD}=-(D-1)\left[A^{\prime\prime}+
 (A^\prime)^2\right]~,\\
 &&R_{\mu D}=0~,
\end{eqnarray}
where prime denotes derivative w.r.t. $y$. Also, ${\widetilde R}_{\mu\nu}$
is the $(D-1)$-dimensional 
Ricci tensor corresponding to the metric ${\widetilde g}_{\mu\nu}$.

{}In the following we will be 
interested in solutions where $\phi$ depends non-trivially on $y$. From the
above equations it then follows that $\phi$ is independent of $x^\mu$.  
The equations of motion for $\phi$ and $A$ then become:
\begin{eqnarray}\label{phi''}
 &&{8\over {D-2}}\left[\phi^{\prime\prime}+(D-1)A^\prime\phi^\prime\right]-
 V_\phi=0~,\\
 \label{phi'A'}
 &&(D-1)(D-2)(A^\prime)^2-{4\over{D-2}}(\phi^\prime)^2+V-
 {{D-1}\over{D-3}}{\widetilde \Lambda}\exp(-2A)=0~,\\
 \label{A''}
 &&(D-2)A^{\prime\prime}+{4\over {D-2}}(\phi^\prime)^2+{1\over {D-3}}
 {\widetilde \Lambda}\exp(-2A)=0~.
\end{eqnarray}
The first equation is the dilaton equation of motion, the second equation
is that for $R_{DD}$, and the third equation is a linear combination of the
latter and the equation for $R_{\mu\nu}$. In fact, the equation 
for $R_{\mu\nu}$
implies that ${\widetilde \Lambda}$ is independent of $x^\mu$ and $y$. In fact,
it is nothing but the cosmological constant of the $(D-1)$-dimensional 
manifold, which is therefore an Einstein manifold, described by the metric 
${\widetilde g}_{\mu\nu}$. Our normalization of ${\widetilde \Lambda}$ is such
that the $(D-1)$-dimensional metric ${\widetilde g}_{\mu\nu}$ satisfies
Einstein's equations
\begin{equation}
 {\widetilde R}_{\mu\nu}-{1\over 2}{\widetilde g}_{\mu\nu}
 {\widetilde R}=-{1\over 2}
{\widetilde g}_{\mu\nu}{\widetilde\Lambda}~,
\end{equation}
so that the $(D-1)$-dimensional Ricci scalar is given by
\begin{equation}
 {\widetilde R}={{D-1}\over{D-3}}{\widetilde\Lambda}~.
\end{equation} 

{}Note that we have only two fields $\phi$ and $A$, yet we have three 
equations. However, only two of these equations are independent. This can
be seen as follows. Using the second equation one can express $\phi^\prime$ 
($A^\prime$) via $A^\prime$ ($\phi^\prime$) and $V$. One can then
compute $\phi^{\prime\prime}$ ($A^{\prime\prime}$) and plug it in the first 
(third) equation. This equation can then be seen to be automatically satisfied
as long as the third (first) equation is satisfied. As usual, this is a
consequence of Bianchi identities.

\section{Solutions with $(D-1)$-dimensional Poincar{\'e} Invariance}

{}In this section we will discuss solutions of the aforementioned equations
with ${\widetilde\Lambda}=0$. In this case the equations of motion read:
\begin{eqnarray}
 &&(D-1)(D-2)(A^\prime)^2-{4\over{D-2}}(\phi^\prime)^2+V=0~,\\
 &&(D-2)A^{\prime\prime}+{4\over {D-2}}(\phi^\prime)^2~=0.
\end{eqnarray}
To study possible solutions to these equations, it is convenient to rewrite
the scalar potential $V$ as follows:
\begin{equation}\label{super}
 V=W_\phi^2-\gamma^2 W^2~,
\end{equation}
where
\begin{equation}
 \gamma={2\sqrt{D-1}\over{D-2}}~,
\end{equation}
and $W$ is some function of $\phi$. Note that this is always possible
(at least for a large class of scalar potentials) if we view (\ref{super})
as a differential equation for $W$. Also note that even if $V$ is a simple
function of $\phi$, generically $W$ is a complicated function, which is due
to the fact that the differential equation (\ref{super}) is highly non-linear.

{}The advantage of using $W$ instead of $V$ is that the equations of motion
for $\phi$ and $A$ can now be written as the first order differential 
equations. Thus, the above two equations are satisfied 
if $\phi$ and $A$ satisfy
\begin{eqnarray}
 &&\phi^\prime=\alpha W_\phi~,\\
 &&A^\prime=\beta W~,
\end{eqnarray}
where
\begin{equation}
 \alpha\equiv\epsilon {\sqrt{D-2}\over 2}~,~~~\beta\equiv-\epsilon
 {2\over (D-2)^{3/2}}~.
\end{equation}
Here $\epsilon=\pm 1$.

{}Here one can ask whether the equations of motion imply that $\phi$ and 
$A$ must satisfy the first order differential equations for some $W$. The
answer to this question is positive. To see this, it is convenient to first
rewrite the equations of motion by treating $A$ as a function of $\phi$
(instead of $y$), while still treating $\phi$ as a function of $y$. Thus, we
have
\begin{eqnarray}
 &&{4\over {D-2}}(\phi^\prime)^2 
 \left[1-{1\over 4}(D-1)(D-2)^2 (A_\phi)^2\right]=V~,\\
 &&A_{\phi\phi} (\phi^\prime)^2=-{{D-2}\over 8} A_\phi V_\phi-
 {1\over {D-2}} V~.
\end{eqnarray}  
Next, let
\begin{equation}
 V\left[1-{1\over 4}(D-1)(D-2)^2 (A_\phi)^2\right]^{-1}\equiv h^2~,
\end{equation}
where $h$ is a function of $\phi$. Then $\phi$ satisfies the first order
differential equation 
\begin{equation}
 \phi^\prime=\alpha h~.
\end{equation}
Now define $W$ via
\begin{equation}
 V=h^2-\gamma^2 W^2~.
\end{equation}
It then follows that $A$ satisfies the first order differential equation
\begin{equation}
 A^\prime=\eta |\beta| W~,
\end{equation}
where $\eta=\pm 1$.
Requiring that the equation for $A^{\prime\prime}$ (or, equivalently, that
for $A_{\phi\phi}$) is satisfied then implies that 
\begin{equation}
 h=-\epsilon\eta W_\phi,
\end{equation}
which is what we wished to show. Note that without loss of generality 
above we have chosen $\eta=-\epsilon$.

\subsection{Non-singularity Conditions}

{}In this subsection we would like to discuss the conditions on $W$ such that
the corresponding solutions do not blow up at finite values of $y$. More
precisely, in this section we will focus on solutions such that
$\phi$ is non-singular\footnote{We will refer to the corresponding domain 
walls as non-singular. However, as we will point out in the following,
some of such solutions are actually singular in the sense that the
$D$-dimensional Ricci scalar $R$ 
blows up, but the singularities are located at $y=\pm\infty$.} 
at finite $y$. (We will discuss singular solutions
in section V.) To begin with note that if $V$ is non-singular,
which we will assume in the following,
then $W$ and $W_\phi$ should (generically) be non-singular as well. This then
guarantees that solutions are continuous for finite values of $\phi$. However,
{\em a priori} it is still possible that $\phi$ blows up at finite values of
$y$. 

{}Note that the question of non-singularity of solutions can be studied
without any reference to the coupling to gravity. Indeed, the equation
we would like to study is
\begin{equation}\label{phi}
 \phi^\prime =\alpha W_\phi~.
\end{equation}
This equation arises in a non-gravitational theory described by the
action
\begin{equation}
 {\cal S}=\int d^D x \left[-{4\over {D-2}}(\partial\phi)^2-{\cal V}(\phi)
 \right]~.
\end{equation}
We can rewrite the potential ${\cal V}$ as
\begin{equation}
 {\cal V}=W_\phi^2
\end{equation}
if we view it as a differential equation for $W$. Note that 
this is possible as long as ${\cal V}$ is non-negative. 
Now let us study solutions
which depend on $y$ only. Then the equation of motion for $\phi$ is given by
(\ref{phi}), which is nothing but the BPS equation. As should be clear from 
our discussion, one can write this equation for any non-negative
potential ${\cal V}$. In
particular, the theory need not be supersymmetric. However, {\em a priori}
there is no guarantee that solutions of this equation, that is, BPS solutions,
satisfying certain physical requirements exist for a given ${\cal V}$. This is
precisely the question we would like to study here.

{}It is convenient to divide the discussion of this question according to
whether $W_\phi$ vanishes for some value(s) of $\phi$ or its non-vanishing
for all real $\phi$. We will first discuss the latter case.

\begin{center}

{\em Non-vanishing $W_\phi$}

\end{center}

{}Suppose $W_\phi$ is non-vanishing for all real $\phi$. In a (globally)
supersymmetric setup this would be equivalent to the statement that 
the theory does not have a supersymmetric vacuum. More generally, we have two
possibilities. The potential ${\cal V}$ has extrema if and only if 
$W_{\phi\phi}$ vanishes for some values of $\phi$. Some of these extrema
may correspond to (local) minima (if ${\cal V}_{\phi\phi}=2
W_\phi W_{\phi\phi\phi}>0$ at the
corresponding values of $\phi$). 
On the other hand, there might be cases where the potential
${\cal V}$ has no extrema at all - this happens if not only $W_\phi$ but also
$W_{\phi\phi}$ is non-vanishing. In this case we have a runaway type of
potential. At any rate, for our discussion here it will have little relevance
whether the potential has local minima or not. As we will point out, 
non-singular BPS solutions exist in either case provided that a certain simple
criterion is satisfied.

{}It might be unfamiliar, and even sound surprising, that potentials with, say,
runaway behavior can have BPS solutions. An example of this type was originally
discussed in \cite{DS}. Before we give our general discussion here, we would
like to give an even simpler example. Thus consider a theory with
\begin{equation}
 W=\rho\phi~,
\end{equation}
where $\rho$ is a parameter. Note that this is a theory with a constant
potential ${\cal V}=\rho^2$. In the globally supersymmetric context 
supersymmetry is completely 
broken in $D$ dimensions. However, there are BPS solutions
with only half of the supersymmetries broken. These are given by
\begin{equation}
 \phi(y)=\alpha\rho y~.
\end{equation}
Thus, even though supersymmetry is broken if we preserve $D$-dimensional 
Lorentz invariance, we can preserve half of the supersymmetries at the cost
of breaking Lorentz invariance, more concretely, to that of a 
$(D-1)$-dimensional theory. If we view this (delocalized) solution as a 
domain wall, then its tension is infinite. In fact, so is the corresponding
central charge in the superalgebra. Thus, for instance, in the context
of ${\cal N}=1$ supersymmetry in $D=4$ 
(in which case we really have one complex
scalar filed in the corresponding chiral supermultiplet) there is a central
extension of the ${\cal N}=1$ superalgebra with an infinite central 
charge\cite{DS}:
\begin{equation}
 \left\{Q_\alpha~,~Q_\beta\right\}=\Sigma_{\alpha\beta}{\cal Z}~,
\end{equation} 
where $\Sigma_{\alpha\beta}$ is proportional to the area tensor in the
plane perpendicular to the $y$ direction, and
\begin{equation}
 {\cal Z}=2\left[W(y=L)-W(y=-L)\right]=({\rm const.})\times L\rightarrow
 \infty~.
\end{equation}
Thus, solutions of this type can be viewed as domain walls with infinite
tension/central charge. As we will see in the following, 
gravitational generalizations of such domain walls, if
they are non-singular, automatically localize gravity. 

{}Next, let us discuss the general condition for such domain walls to be 
non-singular. That is, we would like to find the condition under which
$\phi$ does not blow up at finite values of $y$. For this to be the case,
it is necessary and sufficient that the function 
\begin{equation}\label{non-sing}
 F(\phi)\equiv\int{d\phi\over W_\phi}
\end{equation}
is unbounded at $\phi\rightarrow\pm \infty$. Indeed, let $\phi(y_0)=\phi_0$ be
finite for some finite point $y_0$. Then
\begin{equation}
 \int_{\phi_0}^{\infty}{d\phi\over W_\phi}=\alpha (y_\infty-y_0)~,
\end{equation}
where $y_\infty$ is the point where $\phi$ becomes $\infty$. If the above 
condition is satisfied, then the above integral is divergent, and $y_\infty$ is
either $+\infty$ or $-\infty$ (depending on the form of $W$ and sign of
$\alpha$). Similar considerations apply to $y_{-\infty}$ where $\phi$ becomes
$-\infty$. The above condition can be reformulated directly in terms of $W$:
\begin{center}
 $W$ should not\footnote{More precisely, this is correct up to usual
 ``logarithmic'' factors (that is, $\log(\phi)$, $\log(\log(\phi))$, {\em 
 etc.}, or, more generally, the appropriate products thereof). Thus, for 
 instance, the non-singularity condition on (\ref{non-sing}) 
 is satisfied for $W=\xi\phi^2\log(\phi)$.} 
 grow faster than $\phi^2$ for $\phi\rightarrow\pm \infty$~.
\end{center}
It is clear that there exist infinite choices of $W$ which satisfy this
condition.

{}Let us discuss general properties of such domain walls assuming that the
above non-singularity condition is satisfied. Note that $W_\phi$ never changes
sign as $W_\phi$ is non-vanishing. This then implies that $\phi$ monotonically
increases/decreases from $-\infty/+\infty$ to $+\infty/-\infty$. That is, 
non-singular domain walls of this type have infinite tension (and, 
equivalently, infinite central charge in the supersymmetric context).  

{}Here the following remark is in order. Consider the cases where $W_\phi$
is non-vanishing at finite $\phi$ but goes to zero for $\phi\rightarrow+\infty$
and/or $\phi\rightarrow -\infty$. If $W$ blows up at large $\phi$
(but more slowly than $\sim \phi$, so that $W_\phi$ vanishes there), 
then the tension of the domain wall is 
infinite. However, if $W$ approaches finite values at
$\phi\rightarrow\pm\infty$ (which implies that $W_\phi$ vanishes
at $\phi\rightarrow\pm\infty$), then the tension of the domain wall is finite -
it is proportional to $W(+\infty)-W(-\infty)$, which is finite in this case.
The situation here is similar to that where $W_\phi$ vanishes at finite values
of $\phi$.

\begin{center}
 {\em $W_\phi$ Vanishing at One Point}
\end{center}

{}Let us now discuss the case where $W_\phi$ vanishes for only one value
of $\phi=\phi_0$. This corresponds to a global minimum of the potential.
(More precisely, there might also be a runaway branch with vanishing 
${\cal V}$ in the large $|\phi|$ limit.) 

{}In such cases we can also have non-singular domain walls. First, let us study
the $y$ dependence of $\phi$ near $\phi_0$. We have
\begin{equation}
 W_\phi=W_{\phi\phi}(\phi_0)(\phi-\phi_0)+{1\over 2}W_{\phi\phi\phi}(\phi_0)
 (\phi-\phi_0)^2+\dots~.
\end{equation}
Let us first assume that $W_{\phi\phi}(\phi_0)\not=0$. Then the leading
$y$ dependence of $\phi$ near $\phi_0$ is given by
\begin{equation}
 \phi(y)-\phi_0\sim C\exp(ay)~,
\end{equation}
where $C$ is the integration constant, and $a\equiv \alpha 
W_{\phi\phi}(\phi_0)$. For definiteness let us assume that $a>0$. Then
$\phi$ approaches $\phi_0$ at $y\rightarrow -\infty$. In the complete solution
$\phi$ will then monotonically increase/decrease to $+\infty/-\infty$ at
$y\rightarrow +\infty$ provided that the solution is non-singular. This is the
case provided that $W$ does not grow faster than $\phi^2$ at $\phi\rightarrow
+\infty/-\infty$. If this condition is satisfied only for, say,
$\phi\rightarrow +\infty$, then we only have non-singular solutions with
$\phi\rightarrow\phi_0$ at $y\rightarrow \mp\infty$ and $\phi
\rightarrow +\infty$ at $y\rightarrow\pm\infty$. If it is also satisfied for
$\phi\rightarrow -\infty$, we then also have the corresponding solutions 
with $\phi\rightarrow\phi_0$ at $y\rightarrow \mp\infty$ and $\phi
\rightarrow -\infty$ at $y\rightarrow\pm\infty$.

{}Cases with $W_{\phi\phi}(\phi_0)=0$ can be discussed similarly. In fact,
let us assume that the lowest (non-trivial)
derivative of $W$ which is non-vanishing at
$\phi_0$ is the $k$-th derivative ($k\geq 3$). Then the leading $y$ dependence
of $\phi$ near $\phi_0$ is given by 
\begin{equation}
 (\phi(y)-\phi_0)^{k-2}\sim {a\over y}~,
\end{equation}
where $a\equiv -(k-1)!/(k-2)\alpha W^{(k)}(\phi_0)$. Once again, we have domain
walls interpolating between $\phi=\phi_0$ at $y\rightarrow\mp\infty$ and
$\phi=\pm\infty$ at $y\rightarrow\pm\infty$ provided that the corresponding
non-singularity conditions are satisfied.
 
{}Let us consider a simple example of such domain walls. Let 
\begin{equation}
 W={1\over 2} \zeta \phi^2~.
\end{equation}
Then we have a single minimum at $\phi=0$. Note that the non-singularity
conditions are satisfied for this choice of $W$.
The domain wall solutions are given by
\begin{equation}
 \phi(y)=C\exp(\alpha\zeta y)~,
\end{equation}
where $C$ is an arbitrary integration constant.

\begin{center}
 {\em $W_\phi$ Vanishing at Multiple Points}
\end{center}

{}If $W_\phi$ vanishes at more then one points, call them $\phi_a$, then
we have familiar domain walls (with finite tension) interpolating between
adjacent vacua. However, we can also have domain walls of the aforementioned
type, which interpolate between $\phi={\rm min}(\phi_a)/\phi={\rm max}(\phi_a)$
and $\phi=-\infty/\phi=+\infty$ provided that $W$ does not grow faster then
$\phi^2$ at $\phi\rightarrow -\infty/+\infty$.

{}Before we end this subsection, let us illustrate some of the points discussed
above on the example of the $\phi^4$ theory. More concretely, consider
the potential
\begin{equation}
 {\cal V}=V_0+m^2 \phi^2+\lambda\phi^4~,
\end{equation}
where $\lambda>0$. First consider the case $m^2>0$. In this case we must assume
that $V_0\geq 0$ so that ${\cal V}$ is non-negative. Then the theory has one
minimum, which corresponds to $W_\phi=0$ for $V_0=0$, but for $V_0>0$ 
$W_\phi$ is non-vanishing. In neither case, however, do we have non-singular
domain walls as the non-singularity conditions on $W_\phi$ at $\phi\rightarrow
\pm\infty$ are not satisfied.

{}Next, let us consider the case $m^2<0$. Let 
\begin{equation}
 m^2\equiv -2\lambda v^2~.
\end{equation}
Then we can rewrite the potential as
\begin{equation}
 {\cal V}=(V_0-\lambda v^4)+\lambda (\phi^2-v^2)^2~,
\end{equation}
where we must assume that $V_0\geq \lambda 
v^4$ so that the potential is non-negative
at the minima $\phi=\pm v$. These correspond to $W_\phi=0$ for $V_0=
\lambda v^4$, but
for $V_0>\lambda v^4$ 
$W_\phi$ is non-vanishing. It then follows that for $V_0>\lambda v^4$
there are no non-singular domain walls. For $V_0=\lambda 
v^4$, however, we have the
well-known kink solutions interpolating between the two vacua $\phi=\pm v$.
Note that in this case we do not have non-singular
domain walls interpolating between $\phi=\pm v$ and $\phi\rightarrow\pm\infty$
(the signs are correlated) as the non-singularity conditions are not satisfied.

{}Thus, for the $\phi^4$ theory we have reproduced a known result that 
non-singular domain wall solutions exist (for $m^2<0$)
only if we ``fine-tune'' $V_0$ in
terms of $m^2$ and $\lambda$ as follows:
\begin{equation}
 V_0={m^4\over 4\lambda}~.
\end{equation}
In this particular case this is equivalent to the requirement that
the (two degenerate) ground state(s) of the theory have vanishing vacuum
energy. This is because in this case non-singularity conditions are not
satisfied for non-singular 
infinite tension domain walls to exist. However, as should
be clear from our previous discussions, such domain walls can exist in
theories with ground state(s) with non-vanishing vacuum energy or even
theories with no ground state at all. 

\begin{center}
 {\em Normalizable Modes}
\end{center}

{}Before we end this subsection, we would like to discuss the condition 
for the existence of normalizable modes localized on a domain wall (in the
non-gravitational context discussed in this subsection). Thus, consider
small fluctuations $\varphi(x^\mu,y)$ around the domain wall background. 
The linearized action for $\varphi$ reads (up to surface terms)
\begin{eqnarray}
 {\cal S}[\varphi]\equiv&&{\cal S}[\phi+\varphi]-{\cal S}[\phi]\nonumber\\
 &&=-{1\over \alpha^2}\int d^D x\left\{ 
 \partial_\mu\varphi\partial^\mu\varphi +
 (\varphi^\prime)^2 +\alpha^2
 \left[W_{\phi\phi}^2+W_\phi W_{\phi\phi\phi}\right]\varphi^2\right\}~.
\end{eqnarray}
Here $\phi$ satisfies (\ref{phi}). Let us define
\begin{equation}
 \varphi\equiv \phi^\prime\omega~.
\end{equation}
The action for $\omega$ reads:
\begin{equation}\label{omega}
 {\cal S}[\omega]=-\int d^{D-1}x dy W_\phi^2 \left[
 \partial_\mu\omega\partial^\mu\omega+(\omega^\prime)^2\right]~.
\end{equation}

{}The zero mode is given by the configurations
where $\omega$ is independent of $y$
\begin{equation}
 \omega=\omega(x^\mu)~.
\end{equation}
To have such a normalizable zero mode, the integral
\begin{equation}
 T\equiv 2\int_{-\infty}^{+\infty} dy W_\phi^2=\left. 
 {2\over \alpha} W\right|_{y=-\infty}^{y=+\infty}~,
\end{equation}
which is nothing but the domain wall tension,
must be finite. Thus, domain walls with infinite tension do not have
normalizable zero modes in this context. On the other hand, domain walls with
finite tension do have normalizable zero modes\footnote{As we will see in the
following, once we consider the coupling to gravity, this conclusion is
modified.}. 

{}Here the following remark is in order. The action (\ref{omega}) is actually
exact, that is, the $\omega$ zero mode is a free field. This is due to the fact
that it describes the Goldstone mode of the broken translational invariance
in the $y$ direction.
Thus, let $\phi(y)$ be a solution describing a non-singular domain wall. 
Consider the following fluctuations around this solution: $\phi(y+
\omega(x^\mu))$. Then we have $\phi(y+
\omega(x^\mu))=\phi(y)+\phi^\prime(y)\omega(x^\mu)+{\cal O}(\omega^2)$, so that
$\omega(x^\mu)$ is precisely the linearized 
zero mode described above. However, if we plug
$\phi(y+\omega(x^\mu))$ into the action ${\cal S}[\phi]$, after shifting the
integration variable $y$ (note that integration over $y$ is from $-\infty$ to
$+\infty$ for non-singular domain walls), we obtain the action 
${\cal S}[\omega]$ given in (\ref{omega}) for arbitrary $\omega(x^\mu)$.

{}Also, the equation of motion for $\omega$, which follows from the action 
(\ref{omega}), is given by
\begin{equation}
 W_\phi^2 \partial_\mu\partial^\mu \omega +\left[W_\phi^2 \omega^\prime
 \right]^\prime=0~.
\end{equation}
Let 
\begin{equation}
 \omega(x^\mu,y)={\widetilde \omega}(x^\mu) \sigma(y)~,
\end{equation}
where ${\widetilde\omega}$ satisfies the $(D-1)$-dimensional Klein-Gordon
equation:
\begin{equation}
 \partial_\mu\partial^\mu{\widetilde\omega}=m^2{\widetilde\omega}~.
\end{equation}
Then we have the following equation for $\sigma$:
\begin{equation}
 W_\phi^2 m^2\sigma +\left[W_\phi^2 \sigma^\prime
 \right]^\prime=0~. 
\end{equation}
In particular, a zero mode ($m^2=0$) must satisfy
\begin{equation}\label{sigma'}
 \sigma^\prime=C/W_\phi^2~,
\end{equation}
where $C$ is a constant. Above we claimed that the zero mode corresponds to
taking $C=0$. To see that $C\not=0$ does not give rise to normalizable
zero modes, note that in this case $\sigma$ is a monotonous function of
$y$ as can be seen from (\ref{sigma'}). 
Thus, it either goes to a constant at, say, $y\rightarrow+\infty$, or
is unbounded. In either case, if the domain wall tension is infinite, such
a solution is not normalizable. In fact, it is not difficult to show that
such zero modes are not normalizable even if the domain wall tension is finite.
 
{}Normalizability of other (that is, non-zero) modes can be discussed in a
similar fashion. For this it is sometimes convenient to 
treat $\sigma$ as function of $\phi$. We then have
\begin{equation}
 \left[\sigma_\phi W_\phi^3\right]_\phi +W_\phi m^2\sigma=0~,
\end{equation}
and
\begin{equation}
 ||\varphi||^2\equiv\int dy \varphi^2=
 {1\over\alpha}{\widetilde\omega}^2 \int d\phi W_\phi \sigma^2~.
\end{equation}

{}Thus, for illustrative purposes let us 
consider an example where $W={1\over 2}\zeta\phi^2$. 
As we discussed above, in this case the domain
wall solution is given by
\begin{equation}
 \phi(y)=C\exp(\alpha\zeta y)~,
\end{equation}
where $C$ is an integration constant. The equation for $\sigma$ now reads:
\begin{equation}
 \left[\phi^3\sigma_\phi\right]_\phi +{m^2\over\alpha^2\zeta^2}\phi\sigma=0~.
\end{equation}
The general solution to this equation is given by:
\begin{equation}
 \sigma = D_+ \phi^{-\rho_+} + D_- \phi^{-\rho_-}~,
\end{equation}
where 
\begin{equation} 
 \rho_\pm=1\pm\sqrt{1-{m^2\over\alpha^2\zeta^2}}~,
\end{equation}
and $D_\pm$ are integration constants. Since the domain wall interpolates
between $\phi=0$ and $\phi=\pm\infty$ (depending on the sign of $C$),
it then follows that $||\varphi||^2$ is infinite for all allowed
$m^2$, so that there
are no normalizable modes. Note that the tension of this domain wall is
infinite.

\subsection{What about the Warp Factor?}

{}In this subsection we would like to discuss what happens to the
warp factor once we consider the gravitational generalizations of the
above domain walls. (Note that the discussion of the behavior of $\phi$
is unchanged.) To answer this question, let us look at the equation for
$A$:
\begin{equation}
 A^\prime=\beta W~.
\end{equation}
To begin with let us discuss the behavior of $A$ as a function of $\phi$ 
near the points where $W_\phi$ vanishes. Let us refer to such a point as
$\phi_0$. First, let us focus on the case where $W_{\phi\phi}(\phi_0)\not=0$.
In this case we have 
\begin{equation}
 A^\prime=\beta\left[W(\phi_0)+
 {1\over 2}W_{\phi\phi}(\phi(y)-\phi_0)^2+\dots\right]~.
\end{equation}
Since in this case $\phi(y)-\phi_0\sim C\exp(ay)$ vanishes exponentially fast
for the corresponding limit $y\rightarrow+\infty$ or $y\rightarrow-\infty$,
it then follows that all the contributions to $A$ 
except for that corresponding to
the $W(\phi_0)$ term vanish exponentially as well. That is, $A(y)$ goes to a 
constant if $W(\phi_0)$ vanishes, while in the case $W(\phi_0)\not=0$ the
leading behavior of $A$ as a function of $y$ is given by:
\begin{equation}
 A(y)\sim \beta W(\phi_0) y~.
\end{equation}
This implies that for $W(\phi_0)=0$ the factor $\exp(2A)$ in the metric
goes to a constant in the corresponding limit for $y$, so that the
volume of the compactification in this case is actually infinite. However,
if $W(\phi_0)$ is non-zero and has a correct sign, then $\exp(2A)$ vanishes
exponentially in the corresponding limit for $y$. 

{}In fact, with appropriately
chosen $W$ such that the domain wall interpolates between two adjacent vacua  
we can have the correct behavior at both $y\rightarrow-\infty$ and
$y\rightarrow+\infty$ so that the compactification volume is finite. 
More concretely, let $W_\phi$ vanish at two points, call them $\phi_1$ and
$\phi_2$, and be non-vanishing for $\phi_1<\phi<\phi_2$. Then if $W(\phi_1)$
and $W(\phi_2)$ are non-zero with ${\rm sign}(W(\phi_1))=-{\rm sign}
(W(\phi_2))={\rm sign}
(\beta)$ (that is, $W$ must change sign between $\phi_1$ and $\phi_2$),
$A(y)$ goes to $-\infty$ at both $y\rightarrow\pm\infty$.
Let us
illustrate this with a simple example. Consider the theory with
\begin{equation}
 W=\xi\left[\phi-{1\over 3}\phi^3\right]~.
\end{equation}
The solution for $\phi$ is given by ($y_0$ is the integration constant
corresponding to the center of the domain wall)
\begin{equation}
 \phi(y)=\tanh(\alpha \xi (y-y_0))~,
\end{equation}
while the solution for $A$ reads ($C$ is an integration constant)
\begin{equation}
 A(y)={2\beta\over 3\alpha}\left[\ln(\cosh(\alpha \xi (y-y_0)))-{1\over 4}
 {1\over \cosh^2(\alpha \xi (y-y_0))}\right]+C~.
\end{equation}
We therefore have
\begin{equation}
 \exp(2A)=\exp(2C)
 \left[\cosh(\alpha \xi (y-y_0))\right]^{4\beta/3\alpha}\exp\left[-
 {\beta\over 3
 \alpha}{1\over \cosh^2(\alpha \xi (y-y_0))}\right]~.
\end{equation}
Since $\alpha$ and $\beta$ have opposite signs, it then follows that
$\exp(2A)$ falls off exponentially at large $y$, and the compactification
volume is indeed finite.

{}The cases where not only $W_\phi$ but also higher derivatives vanish 
at $\phi_0$ can be discussed similarly. Suppose the lowest non-trivial
derivative of $W$ which is non-vanishing is the $k$-th derivative ($k\geq 3$).
Then we have
\begin{equation}
 W=W(\phi_0)+{1\over k!}W^{(k)}(\phi_0)(\phi-\phi_0)^k+\dots~.
\end{equation}
Since in this case $(\phi(y)-\phi_0)^{k-2}\sim a/y$, then it follows that
in the corresponding limit for $y$ the contributions to $A$ from all the terms
except the $W(\phi_0)$ term vanish. For instance, the first subleading term
proportional to $W^{(k)}(\phi_0)$ goes to zero as $1/y^{2/(k-2)}$. Thus, as 
in the previous case, here we have a finite volume compactification only if
$W(\phi_0)\not=0$.

{}Let us now discuss the behavior of $A$ in the cases where in the 
corresponding limit for $y$ $\phi$ goes to $\pm \infty$. (This is the case
of domain walls with infinite tension.) Let us assume that for large $\phi$
$W$ has the following behavior:
\begin{equation}
 W\sim \xi \phi^\nu~,
\end{equation}
where $\nu\leq 2$ so that the non-singularity conditions are satisfied.
Moreover, first let us consider the case where $\nu>0$ as the analog of the
$\nu<0$ case requires a separate treatment.
Let us discuss the $\nu=2$ and $\nu<2$ cases separately. 

{}For $\nu=2$ we have $W_\phi\sim 2\xi\phi$, and
\begin{equation}
 \phi(y)\sim C\exp(2\alpha\xi y)~,
\end{equation}
where C is the integration constant. In this case we therefore have the 
following leading behavior for $A$:
\begin{equation}
 A(y)\sim {\beta \over 4\alpha} C^2 \exp(4\alpha\xi y)\sim 
 {\beta \over 4\alpha} \phi^2~,
\end{equation}
so that $A(y)$ goes to $-\infty$ exponentially fast (recall that $\alpha$ 
and $\beta$ have opposite signs), and $\exp(2A)$ goes to zero very fast,
more concretely, as an exponential of an exponential.

{}Next, let us discuss $0<\nu<2$ cases. We then have $W_\phi\sim \nu\xi 
\phi^{\nu-1}$, and
\begin{equation}
 \phi(y)\sim \left[\alpha\nu(2-\nu)\xi\right]^{1\over {2-\nu}}
 y^{1\over{2-\nu}}~.
\end{equation}
In this case we therefore have the following leading behavior for $A$:
\begin{equation}
 A(y)\sim {\beta \over 2\nu \alpha} \left[\alpha\nu(2-\nu)\xi\right]^{2\over 
 {2-\nu}}
 y^{2\over{2-\nu}}\sim {\beta \over 2\nu \alpha}\phi^2~,
\end{equation}
so for $0<\nu<2$ $A(y)$ goes to $-\infty$ as a power of $y$. Let
\begin{equation}
 u\equiv \left|{\beta \over \nu \alpha} 
 \left[\alpha\nu(2-\nu)\xi\right]^{2\over 
 {2-\nu}}\right|~.
\end{equation}
Then we have 
\begin{equation}\label{warpinf}
 \exp(2A)\sim \exp(-u |y|^{2\over{2-\nu}})~.
\end{equation}
Thus, $\exp(2A)$ rapidly decays to zero. In particular,
for $\nu=1$ we have a Gaussian fall-off for $\exp(2A)$, while
$\phi$ is linear 
in $y$ for large $y$. So, the compactification volume is finite
as long as $0<\nu\leq 2$.

{}Thus, as we see, such 
infinite tension domain walls localize gravity even more
efficiently than usual domain walls interpolating between adjacent minima.
In the latter case the warp factor $\exp(2A)$ decays as
\begin{equation}
 \exp(2A)\sim\exp(-{\widetilde u}|y|)~,
\end{equation}
where ${\widetilde u}\equiv|2\beta W(\phi_0)|$. On the other hand, for
$0<\nu<2$ the exponent of $|y|$ in (\ref{warpinf}) is larger than 1, while
for $\nu=2$ $\exp(2A)$ decays even more rapidly, namely, as an exponential
of an exponential.

{}Next, let us assume that for large $\phi$ $W$ has the following behavior:
\begin{equation}
 W\sim \zeta+\xi\phi^{-\sigma}~.
\end{equation}
where $\sigma>0$.
Here we have written the leading constant term $\zeta$ which $W$ approaches
in the infinite $\phi$ limit, and also the next-to-leading term which vanishes
in this limit. We then have $W_\phi\sim -\sigma \xi \phi^{-\sigma-1}$, and
\begin{equation}
 \phi(y)\sim \left[-\alpha\sigma(2+\sigma)\xi\right]^{1\over {2+\sigma}}
 y^{1\over{2+\sigma}}~.
\end{equation}
In this case we therefore have the following leading behavior for $A$:
\begin{equation}
 A(y)\sim \beta\zeta y -{\beta\over 2\sigma\alpha}
 \left[-\alpha\sigma(2+\sigma)\xi y\right]^{2\over{2+\sigma}}~,
\end{equation}
so that for $\zeta\not=0$ the leading behavior is
\begin{equation}
 A(y)\sim\beta\zeta y~,
\end{equation}
while for $\zeta=0$ we have
\begin{equation}
 A(y)\sim -{\beta\over 2\sigma\alpha}\phi^2~.
\end{equation}
Since the signs of $\alpha$ and $\beta$ are opposite, in the
$\zeta=0$ case $A$ goes to $+\infty$ (instead of $-\infty$), so that 
$\exp(2A)$ blows up, and the compactification volume is infinite. However,
for non-zero $\zeta$ with the appropriate sign $A$ goes to $-\infty$, so that
$\exp(2A)$ decays exponentially, and the compactification volume is finite
(provided that $\exp(2A)$ goes to zero fast enough 
in the opposite limit for $y$ as well).
Here the situation is analogous to what we have found for the behavior of $A$
near the points where $W_\phi$ vanishes. In particular, to have decaying
$\exp(2A)$ 
we need a constant term $\zeta$ in $W$ in the corresponding
limit.

{}Before we end this subsection, let us discuss the analog of the $\nu=0$
case above. Thus, let us assume that for large $\phi$ $W$ has the following 
behavior:
\begin{equation}
 W\sim \xi\ln(\phi)~.
\end{equation}
In this case we have $W_\phi\sim\xi/\phi$, and
\begin{equation}
 \phi(y)\sim(2\alpha\xi)^{1\over 2} y^{1\over 2}~.
\end{equation}
For definiteness let as assume that $\alpha\xi>0$, so that we are dealing 
with the $y\rightarrow +\infty$ limit here.
The warp factor then has the following leading behavior:
\begin{equation}
 A(y)\sim{1\over 2 }\beta\xi y\ln(y)~.
\end{equation}
Note that $\beta\xi<0$ (which follows from the fact that $\alpha\xi>0$ 
and $\alpha\beta<0$), which implies that in the $y\rightarrow +\infty$ limit
$A(y)$ goes to $-\infty$, and the compactification volume is finite
(once again, provided that $\exp(2A)$ goes to zero fast enough in the $y
\rightarrow -\infty$ limit).

\section{Killing Spinors and Supersymmetry}

{}In this section we would like to discuss the aforementioned domain walls in
the supersymmetric context, where their BPS property is expected to persist
to the quantum level. In particular, such domain walls, as usual, are
expected to be stable.

{}The domain walls we study in this paper preserve one half of the
original supersymmetries in $D$ dimensions. More precisely, the domain walls
preserve one half of the supersymmetries of a supersymmetric minimum of the
scalar potential $V$ if such minima are present. 
However, one of the key 
points is that domain walls have the same number of supersymmetries (which
act trivially on the domain wall) even if $V$ has no supersymmetric minima at
all. In this sense such domain walls are similar to those in the previous case.

{}To see the aforementioned properties of domain walls explicitly, let us
study the corresponding Killing spinor equations, which (up to equivalent
representations) read:
\begin{eqnarray}\label{Killing1}
 &&\left[\Gamma^M\partial_M\phi-\alpha W_\phi\right]\varepsilon=0~,\\
 &&\left[{\cal D}_M-{1\over 2}\beta W \Gamma_M\right] \varepsilon=0~.
 \label{Killing2}
\end{eqnarray}
Here $\varepsilon$ is a Killing spinor, $\Gamma_M$ are $D$-dimensional
Dirac gamma matrices satisfying
\begin{equation}
 \left\{\Gamma_M~,~\Gamma_N\right\}=2 G_{MN}~,
\end{equation}
and ${\cal D}_M$ is the covariant derivative
\begin{equation}
 {\cal D}_M\equiv \partial_M +{1\over 4}\Gamma_{AB}{\omega^{AB}}_M~.
\end{equation}
The spin connection ${\omega^{AB}}_M$ is defined via the vielbeins
${e^A}_M$ in the usual way (here the capital Latin indices $A,B,\dots=1,\dots,
D$ are lowered and raised with the $D$-dimensional Minkowski metric
$\eta_{AB}$ and its inverse, while
the capital Latin indices $M,N,\dots=1,\dots,D$ are lowered and raised with
the $D$-dimensional metric $G_{MN}$ and its inverse). Furthermore,
\begin{equation}
 \Gamma_{AB}\equiv{1\over 2}\left[\Gamma_A~,~\Gamma_B\right]~,
\end{equation} 
where $\Gamma_A$ are the constant Dirac gamma matrices satisfying
\begin{equation}
 \left\{\Gamma_A~,~\Gamma_B\right\}=2 \eta_{AB}~.
\end{equation}
Finally, $W$, which determines the scalar potential via (\ref{super}), is
interpreted as the superpotential in this context.

{}Note that (\ref{Killing1}) comes from the requirement that the variation
of the superpartner $\lambda$ of the scalar field $\phi$ vanish. On the
other hand, (\ref{Killing2}) arises from the requirement that the gravitino
$\psi_M$ have vanishing variations under the corresponding supersymmetry 
transformations.

{}Next, we would like to study the above Killing spinor equations in the
warped backgrounds of the form (\ref{warped}) with the 
flat $(D-1)$-dimensional metric ${\widetilde g}_{\mu\nu}=\eta_{\mu\nu}$
(and the scalar field depending only on $y$):
\begin{eqnarray}\label{Kill1}
 &&\left[\Gamma_D\phi^\prime-\alpha W_\phi\right]\varepsilon=0~,\\
 \label{Kill2}
 &&\varepsilon^\prime-{1\over 2}\beta W\Gamma_D\varepsilon=0~,\\
 \label{Kill3}
 &&\partial_\mu \varepsilon+{1\over 2}\exp(A)
 {\widetilde \Gamma}_\mu\left[A^\prime\Gamma_D-
 \beta W\right]\varepsilon=0~.
\end{eqnarray}
Here ${\widetilde \Gamma}_\mu$ are the constant $(D-1)$-dimensional Dirac gamma
matrices satisfying
\begin{equation}
 \left\{{\widetilde \Gamma}_\mu~,~{\widetilde \Gamma}_\nu\right\}=
 2{\widetilde g}_{\mu\nu}=2\eta_{\mu\nu}~.
\end{equation}
Also, note that $\Gamma_D$, which is the $D$-dimensional Dirac
gamma matrix $\Gamma_M$ with $M=D$ (that is, the Dirac gamma matrix
corresponding to the $x^D=y$ direction) is constant in this background.

{}Note that (\ref{Kill1}), which is an algebraic equation for $\varepsilon$, 
comes from (\ref{Killing1}), while the two differential equations 
(\ref{Kill2}) and (\ref{Kill3}) come from 
(\ref{Killing2}). These last two equations correspond to the coupling to 
gravity, while (\ref{Kill1}) appears already in a non-gravitational system
with the scalar potential ${\cal V}=W_\phi^2$. 

{}Since on the solution we have $\phi^\prime=\alpha W_\phi$ and
$A^\prime=\beta W$, the Killing spinor 
satisfying (\ref{Kill1}), (\ref{Kill2}) and (\ref{Kill3}) 
has positive helicity w.r.t. $\Gamma_D$, and is given by
\begin{equation}\label{Kill+}
 \varepsilon_+=\exp(A/2)\varepsilon_+^{(0)}~,
\end{equation}
where $\varepsilon_+^{(0)}$ is a {\em constant} spinor of positive helicity.

{}Actually, the above conclusion that we only have Killing spinors
of positive helicity is only correct if $W$ has non-trivial $\phi$ dependence.
Thus, consider the case where $W$ is constant. It is clear that 
(\ref{Kill1}) is then satisfied for both positive as well as negative
helicity spinors (as long as $\phi$ is constant). As to (\ref{Kill2}) and
(\ref{Kill3}), the general solution to this set of equations for
constant $W$ is given by
\begin{equation}
 \varepsilon=\left[\beta W\exp(A/2) x^\mu{\widetilde \Gamma}_\mu+
 \exp(-A/2)\right]\varepsilon_-^{(0)}+\exp(A/2)\varepsilon_+^{(0)}~,
\end{equation}
where $\varepsilon_-^{(0)}$ and $\varepsilon_+^{(0)}$ are {\em constant}
negative respectively positive helicity spinors. Thus, for constant $W$, that
is, in the AdS space (for constant $W$ we have constant $V=-\gamma^2 W^2<0$)
we have Killing spinors of both positive and negative helicity. However,
if we consider a domain wall for a non-constant superpotential $W$, then
we have only the positive helicity Killing spinor (\ref{Kill+}). 
Thus, such
domain walls are BPS objects preserving one half of the 
supersymmetries.

{}The BPS property of such a domain wall implies that the modes on the wall 
have boson-fermion degeneracy corresponding to the unbroken 
supersymmetries\footnote{Actually, the state corresponding to the domain wall 
itself does not have a fermionic superpartner as the unbroken supersymmetry
acts trivially on it.}. Thus, for instance, we expect to find a normalizable 
gravitino zero mode localized on the wall. In fact, this zero mode is
given by the variation of $\psi_\mu$ corresponding to the supersymmetry 
transformation parameter $\epsilon$ for which the variations of
$\lambda$ and $\psi_D$ vanish. Such a spinor is given by
\begin{equation}
 \varepsilon=\exp(A/2)\chi_+~,
\end{equation}
where $\chi_+=\chi_+(x^\mu)$ is an arbitrary spinor of positive helicity 
(w.r.t. $\Gamma_D$) which depends on $x^\mu$ but is independent of $y$. The 
corresponding gravitino
variation is given by
\begin{equation}
 \delta\psi_\mu=\exp(A/2)\partial_\mu\chi_+~,
\end{equation}
and it is not difficult to check that $\delta\psi_\mu$ defined this way 
has precisely the correct
normalization (in terms of powers of the warp factor) for it to be the
superpartner of the four dimensional graviton. In fact, both the Einstein
and the Rarita-Schwinger terms in the $(D-1)$-dimensional action are
proportional to 
\begin{equation}\label{Planck}
 M_P^{D-2}\int_{-\infty}^{+\infty} dy \exp\left[(D-3)A\right]\equiv {\widetilde
 M}_P^{D-3}~,
\end{equation}
where $M_P$ is the $D$-dimensional Planck scale, while ${\widetilde M}_P$ is 
the $(D-1)$-dimensional Planck scale. Thus, supersymmetric BPS domain walls
discussed in this paper localize supergravity.

{}Before we end this section, we would like to make the following comment.
As we have already mentioned, in the supersymmetric context BPS domain walls 
are expected to be stable. This is the case not only for the domain walls
interpolating
between two supersymmetric AdS minima, but also for the infinite
tension domain walls (or, more precisely, their gravitational counterparts).
That is, the $D$-dimensional scalar potential $V$ might have no minima at all,
yet the corresponding BPS domain walls are supersymmetric and stable.
In particular, in this context there cannot be any tachyonic modes present, 
which is guaranteed by the unbroken supersymmetries. 

{}Finally, the following remark is in order. The above discussion applies
only to the cases where we can sypersymmetrize the system of the scalar field
$\phi$ coupled to gravity. {\em A priori} such a supersymmetrization might not
exist for a given potential $V(\phi)$. In particular, it might not be possible
to couple the aforementioned bosonic system to fermions in a fashion consistent
with the corresponding superalgebra. However, the question of whether a given
potential can be obtained from a supergravity theory is outside of the scope of
this paper.

\section{Solutions with ${\widetilde \Lambda}\not=0$}

{}In this section we will discuss solutions with non-zero $(D-1)$-dimensional
cosmological constant ${\widetilde \Lambda}$. Thus, let us look at the
equations for $A$ and $\phi$, namely, (\ref{phi'A'}) and (\ref{A''}).
In particular, let us rewrite (\ref{A''})
as follows:
\begin{equation}\label{eq}
 {4\over{D-2}}(\phi^\prime)^2\exp(2A)+{1\over{D-3}}{\widetilde\Lambda}=
 -(D-2)A^{\prime\prime}\exp(2A)~. 
\end{equation}
{}To begin with, let us discuss non-singular domain walls.
To have a non-singular domain wall with finite
${\widetilde M}_P$, it is necessary that $\exp(A)$ asymptotically goes to 
zero. It then follows that $A^{\prime\prime}\exp(2A)$ (as well as,
say, $(A^{\prime})^2
\exp(2A)$) asymptotically goes to zero as well. Then 
the r.h.s. of (\ref{eq}) asymptotically goes to zero at $y\rightarrow \pm
\infty$, while the l.h.s. is strictly positive definite for
${\widetilde\Lambda}>0$. This then implies that ${\widetilde\Lambda}$ cannot 
be positive for non-singular domain walls that localize gravity.  

\subsection{Non-singular Domain Walls with ${\widetilde \Lambda}<0$}

{}Here we can ask whether there exist non-singular domain walls with 
negative cosmological constant. The answer to this question is affirmative.
To construct such a domain wall, pick $A(y)$ such that ${\widetilde M}_P$ is
finite. That is, at $y\rightarrow\pm\infty$ 
$\exp[(D-3)A]$ goes to zero faster than $1/y$.
Moreover, let $A(y)$ be such that
$A^{\prime\prime}$ is non-positive everywhere. This then guarantees that the
r.h.s. of the equation (which follows from (\ref{A''}))
\begin{equation}\label{EQ}
 {4\over{D-2}}(\phi^\prime)^2=-{1\over{D-3}}{\widetilde\Lambda}\exp(-2A)
 -(D-2)A^{\prime\prime}
\end{equation} 
is strictly positive. We can then solve for $\phi(y)$, and express $y$ in terms
of $\phi$ as $\phi(y)$ is a monotonous function of $y$. Using (\ref{phi'A'})
we can express the potential $V$ as a function of $y$, and, subsequently, as
a function of $\phi$. This way we can construct an infinite class of 
non-singular domain walls with ${\widetilde \Lambda}<0$ which localize 
gravity\footnote{A similar procedure was described in \cite{Freedman}. However,
in \cite{Freedman} the second order equations of motion for $\phi$ and $A$ were
rewritten as the first order equations. Unlike the ${\widetilde \Lambda}=0$ 
case, for non-zero (in particular, negative) ${\widetilde\Lambda}$ 
in this process one loses a class of solutions. Thus, it is not difficult to 
show that, for any choice of $V$, 
the first order equations of \cite{Freedman} do not have non-singular
solutions with ${\widetilde\Lambda}<0$ which localize gravity. 
More concretely, in such solutions $A^\prime$ must change sign for some
finite $y$ as $A$ goes to $-\infty$ for $y\rightarrow\pm\infty$. However,
solutions of this type are lost in the process of taking the
square root while deriving a first order equation for $A^\prime$ from 
(\ref{phi'A'}) - see below.}.    

{}Here we would like to study some properties of such domain walls. In
particular, let us understand the asymptotic behavior of $\phi$ and $A$ at
large $y$. From (\ref{EQ}) we obtain the following leading behavior for
$\phi^\prime$ at large $y$:
\begin{equation}
 {4\over{D-2}}(\phi^\prime)^2\sim-{1\over{D-3}}{\widetilde{\Lambda}}
 \exp(-2A)~.
\end{equation}
On the other hand, from (\ref{phi'A'}) we have:
\begin{equation}
 V\sim{{D-2}\over{D-3}}{\widetilde{\Lambda}}
 \exp(-2A)~.
\end{equation}
That is, at large $y$ the potential $V$ behaves as
\begin{equation}
 V\sim-4(\phi^\prime)^2~,
\end{equation}
so that it is unbounded below there. This, in particular, implies that
such solutions with negative ${\widetilde\Lambda}$ cannot asymptotically
approach AdS space (that is, they cannot be of the kink type).

{}Let us define $\chi(\phi)$ at large $\phi$
via 
\begin{equation}
 V(\phi)\sim -4[\chi(\phi)]^2~.
\end{equation}
Then at large $y$ we have
\begin{equation}
 \phi^\prime\sim \chi(\phi)~.
\end{equation}
Note that, for the domain wall to be non-singular, $\chi(\phi)$ must not 
(up to usual ``logarithmic'' factors - see below) grow faster than $\sim\phi$. 

{}Thus, let $\chi(\phi)\sim \xi\phi^\nu$, where $0<\nu<1$. Then we have 
\begin{equation}
 \phi(y)\sim[(1-\nu)\xi y]^{1\over{1-\nu}}~,
\end{equation}
and 
\begin{equation}
 A(y)\sim-{\nu\over{1-\nu}}\ln(\xi y)~.
\end{equation}
Note that to have finite ${\widetilde M}_P$ we must assume that $\exp[(D-3)A]$
goes to zero at large $y$ faster than $1/y$. This then implies the 
following restriction on $\nu$:
\begin{equation}
 \nu>{1\over{D-2}}~.
\end{equation} 
Thus, if this condition is not satisfied, then the corresponding domain walls 
with negative ${\widetilde\Lambda}$ do not localize gravity. This is to
be contrasted with the ${\widetilde\Lambda}=0$ case where such potentials
do give rise to walls which localize gravity. Also, note that to have finite
volume in the $y$ direction, which is given by
\begin{equation}
 {\rm Vol}=\int dy \exp[(D-1)A]~,
\end{equation}
we must assume that $\exp[(D-1)A]$ goes to zero at large $y$ faster
than $1/y$. This is satisfied provided that $\nu>1/D$. Thus, for
$1/D<\nu<1/(D-2)$ the compactification volume is finite, yet gravity is
{\em not} localized.

{}Next, assume that $\chi(\phi)\sim \phi\rho(\ln|\phi|)$. For the domain wall
to be non-singular, the function $\rho(u)$ must be such that the integral
\begin{equation}
 \int{du\over\rho(u)}
\end{equation}
is unbounded at large $u$. The leading behavior for $A$ is then
obtained from the following equation:
\begin{equation}
 A^\prime=-\rho(-A)~.
\end{equation}
Thus, for instance, if $\rho(u)\sim\zeta u^\sigma$, where $0\leq\sigma<1$, 
then 
\begin{equation}
 A\sim-[(1-\sigma)\zeta y]^{1\over{1-\sigma}}~.
\end{equation}
If $\rho(u)\sim\zeta u$, then 
\begin{equation}
 A\sim C\exp(\zeta y)~.
\end{equation}
Here $C$ is a negative integration constant, and $\zeta y$ is assumed to be
positive.

\begin{center}
 {\em First Order Equations}
\end{center}

{}Sometimes it is convenient to rewrite the second order equations of motion
for $\phi$ and $A$ as the first order equations. However, as we have already 
mentioned, for non-zero (in particular, negative) ${\widetilde\Lambda}$ some
extra care is needed not to lose some of the solutions.

{}Thus, consider the case of non-singular domain walls with ${\widetilde
\Lambda}<0$ such that $A$ goes to $-\infty$ at $y\rightarrow\pm \infty$.
Then it follows that at some finite $y$, call it $y_0$, $A^\prime$ must 
change sign. It is then not difficult to show that the first order
equations for $\phi$ and $A$ are given by:
\begin{eqnarray}
 &&\phi^\prime=\alpha h~,\\
 &&A^\prime={\rm sign}(y-y_0)\beta\Omega W~,
\end{eqnarray}
where
\begin{eqnarray}
 &&W_\phi={\rm sign}(y-y_0)\Omega h~,\\
 &&\Omega=\sqrt{1+{{D-1}\over{D-3}}{{\widetilde\Lambda}\over\gamma^2 W^2}
 \exp(-2A)}~,\\
 &&V=h^2-\gamma^2 W^2~.
\end{eqnarray}
Here we note that $W$ does not vanish anywhere, and has an extremum at
$y=y_0$. Also, $\phi^\prime$ does not vanish at $y=y_0$, where by assumption
$A^{\prime\prime}$ is non-positive.

\subsection{Singular Domain Walls}

{}By a singular domain wall we mean a solution where $\phi$ blows up at some
finite value(s) of $y$. We can then have domain walls that are singular on
both sides (they interpolate between $y_-$ and $y_+$, $y_\pm$ being the
points where $\phi$ blows up), or domain walls that are singular on one side 
only (that is, those interpolating between $-\infty$ and $y_+$, or
$y_-$ and $+\infty$). To begin with, note that if $\phi$ blows up at some
finite $y$, call it $y_*$, then $A$ is bounded above at $y=y_*$. Indeed,
let us rewrite (\ref{eq}) as follows:
\begin{equation}\label{A''V}
 \exp(2A)\left[{4\over{D-2}}(\phi^\prime)^2+
 (D-2)A^{\prime\prime}\right]= -{1\over{D-3}}{\widetilde\Lambda}~.
\end{equation}
If $A$ goes to $+\infty$ at $y=y_*$, then $A^{\prime\prime}$
also goes to $+\infty$, and the above equation cannot be satisfied. 

{}Thus, $A$ goes to either $-\infty$ or a constant at $y=y_*$. 
In either case $A^{\prime\prime}$ goes to $-\infty$ there (which follows
from (\ref{A''V})). This then implies that for domain walls singular
on both sides the compactification volume is automatically finite if
we cut off the space in the $y$ direction at the singularity\footnote{Here
one might suspect that not all singularities are physical. The purpose 
of the next subsection is to determine the conditions for physically allowed 
singularities.}. Similarly,
for domain walls singular on one side only the compactification volume is
finite as long as on the non-singular side $\exp(2A)$ goes to zero
fast enough.

{}We now wish to
show that if the scalar potential $V$ is bounded above at the singularity, 
then $A$ necessarily goes to $-\infty$ there (that is, it cannot go to 
a constant). To see this, consider the sum of (\ref{phi'A'}) and (\ref{A''}):
\begin{equation}\label{A'A''}
 (D-2)\left[A^{\prime\prime}+(D-1)(A^\prime)^2\right]+V-
 {{D-2}\over{D-3}}{\widetilde\Lambda}\exp(-2A)=0~.
\end{equation} 
Now suppose $V$ is bounded above at $y=y_*$. Let us assume that $A$ goes
to a constant at $y=y_*$. It then follows from (\ref{A'A''}) that 
$(A^\prime)^2$
goes to $\infty$. In fact, $(A^\prime)^2$
goes to $\infty$ at least as fast as $-A^{\prime\prime}$. However, from
(\ref{A''V}) it follows that $-A^{\prime\prime}$ goes to $\infty$ as fast as
$(\phi^\prime)^2$. We therefore conclude that $A^\prime$ blows up at least
as fast as $\phi^\prime$. This, however, implies that $A$ cannot go to a 
constant at $y=y_*$ as $\phi$ blows up there.

{}Here we note that if we relax the requirement that $V$ be bounded above at 
singularities, then $A$ could
{\em a priori}
go to a constant at a singularity. However, as we will point out 
in the next subsection, for singularities to be physical 
an additional non-trivial condition must be satisfied. 
As we will see in a moment,
singular domain walls with the property that $A$ goes to a constant
at at least one singularity do not satisfy this consistency condition. 

\subsection{Physically Allowed Singularities}  

{}Let us substitute the domain wall ans{\"a}tz into the action $S$ given
in (\ref{action}). We then obtain (here $({\widetilde\nabla}\phi)^2\equiv
{\widetilde g}^{\mu\nu}{\widetilde \nabla}_\mu\phi{\widetilde \nabla}_\nu
\phi$):
\begin{equation}
 S/M_P^{D-2}=\int d^D x\sqrt{-{\widetilde g}}\exp[(D-3)A]\left[{\widetilde R}-
 {1\over\alpha^2}({\widetilde \nabla}\phi)^2\right]-
 \int d^{D-1}x \sqrt{-{\widetilde g}} E[A,\phi]~, 
\end{equation}
where the energy functional $E[A,\phi]$ is given by
\begin{equation}
 E[A,\phi]=\int dy \exp[(D-1)A]\left[(D-1)[2A^{\prime\prime}+D(A^\prime)^2]
 +{1\over \alpha^2}(\phi^\prime)^2 +V(\phi)\right]~.
\end{equation}
Here the integral over $y$ is taken from $y_-$ to $y_+$, where $y_\pm$ 
correspond to the edges of the domain wall ($y_\pm$ can be finite or
infinite). This energy functional can be rewritten as
\begin{eqnarray}
 E[A,\phi]=&&\int dy \exp[(D-1)A]\left[
 {1\over \alpha^2}(\phi^\prime)^2 -
 (D-1)(D-2)(A^\prime)^2
 +V(\phi)\right]+\nonumber\\
 \label{energy}
 &&2(D-1)\left[A^\prime\exp[(D-1)A]\right]
 \Big|^{y_+}_{y_-}~. 
\end{eqnarray}
The boundary term will become important in the following, so we will keep it.
Here we note that (\ref{phi''}) and a linear combination of (\ref{phi'A'}) and
(\ref{A''}) with
${\widetilde \Lambda}=0$ are the Euler-Lagrange equations for
the energy functional (\ref{energy}) \cite{Skenderis}. 
Similarly, for
non-zero ${\widetilde \Lambda}$ the same equations 
are the Euler-Lagrange equations for
the functional
\begin{equation}
 {\widehat E}[A,\phi]\equiv
 E[A,\phi]-\int dy {{D-1}\over{D-3}} {\widetilde\Lambda} \exp[(D-3)A]~.
\end{equation}

{}Using the equations of motion for $\phi$ and $A$, it is not difficult to
show that on the solution we have
\begin{eqnarray}
 E[A,\phi]={\widetilde\Lambda} \int dy \exp[(D-3)A]+ 
 2\left[A^\prime\exp[(D-1)A]\right]
 \Big|^{y_+}_{y_-}~. 
\end{eqnarray}
This then implies that on the solution, where ${\widetilde\nabla}_\mu
\phi=0$, and ${\widetilde g}_{\mu\nu}$ is independent of $y$, we have  
\begin{equation}\label{SSS}
 S={\widetilde M}_P^{D-3}\int d^{D-1}x \sqrt{-{\widetilde g}}
 \left[{\widetilde R}-{\widetilde \Lambda}\right]
 -2M_P^{D-2}
 \left[A^\prime\exp[(D-1)A]\right]
 \Big|^{y_+}_{y_-} \int d^{D-1} x\sqrt{-{\widetilde g}}~,
\end{equation}
where 
\begin{equation}
 {\widetilde M}_P^{D-3}\equiv M_P^{D-2} \int_{y_-}^{y_+} dy \exp[(D-3)A]~.
\end{equation}
This coincides with the $(D-1)$-dimensional action with the Planck scale
${\widetilde M}_P$ and the cosmological constant ${\widetilde \Lambda}$
\begin{equation}
 {\widetilde S}={\widetilde M}_P^{D-3}\int d^{D-1}x\sqrt{-{\widetilde g}}
 \left[{\widetilde R}-{\widetilde \Lambda}\right]
\end{equation}
up to the surface term
\begin{equation}
 -{2\over{D-1}}M_P^{D-2} {\cal A}^\prime\Big|_{y_-}^{y_+}\int d^{D-1} x
 \sqrt{-{\widetilde g}}~,
\end{equation}
where 
\begin{equation}
 {\cal A}\equiv\exp[(D-1)A]~.
\end{equation}
Thus, to have a consistent $(D-1)$-dimensional interpretation, we must require
that the aforementioned boundary contribution vanish. That is, we have
an additional consistency condition:
\begin{equation}\label{allowed}
 {\cal A}^\prime\Big|_{y_-}^{y_+}=0~.
\end{equation}
Note that for non-singular domain walls with finite ${\widetilde M}_P$
this condition is automatically 
satisfied - indeed, for such walls ${\cal A}$ and, therefore, ${\cal A}^\prime$
asymptotically go to zero at $y_\pm=\pm\infty$. However, for singular domain
walls this condition is non-trivial. In fact, in the case of singular domain
walls ${\cal A}^\prime(y_\pm)$ need not even be finite. For the condition 
(\ref{allowed}) to be meaningful, we must then require that  
\begin{equation}\label{allowed0}
 {\cal A}^\prime(y_\pm)~{\mbox{are finite, and}}~{\cal A}^\prime(y_+)-
 {\cal A}^\prime(y_-)=0~.
\end{equation}

{}Note that if ${\cal A}$ goes to zero at singularities, then  
${\cal A}^\prime$ is non-positive at $y=y_+$, and
non-negative at $y=y_-$, so that the condition (\ref{allowed}) is satisfied
if and only if
\begin{equation}\label{allowed1}
 {\cal A}^\prime(y_\pm)=0~.
\end{equation} 
As we will show in a moment, singularities where ${\cal A}$ is finite do
not satisfy the condition (\ref{allowed0}). This then implies that at
physically allowed singularities ${\cal A}$ as well ${\cal A}^\prime$
must necessarily vanish.

\begin{center}
 {\em Potentials Bounded Above}
\end{center}

{}Let us first consider
the case where $V$ is bounded above at singularities. In this case
${\cal A}$ goes
to zero at $y_\pm$, and, therefore, 
${\cal A}^\prime$ must also go to zero there. 

{}If ${\widetilde \Lambda}\leq 0$ 
these conditions imply a certain restriction on the
scalar potential. Thus, multiplying both sides of (\ref{A'A''})
by $\exp[(D-1)A]$, we obtain
\begin{equation}\label{A''V1}
 V=-{{D-2}\over{D-1}}{{\cal A}^{\prime\prime}\over {\cal A}}+
 {{D-2}\over{D-3}}{{\widetilde\Lambda}\over {\cal A}^{2\over{D-1}}}~.
\end{equation} 
Now,  
both ${\cal A}$ and ${\cal A}^\prime$ must vanish at $y_\pm$. Then it follows
that ${\cal A}^{\prime\prime}$ is non-negative at $y_\pm$. This then implies 
that for ${\widetilde \Lambda}\leq 0$ 
at the singularities the scalar potential must be unbounded 
below\footnote{This is a much stronger condition than that recently
conjectured in \cite{gubser} to be necessary for a singularity to be 
physical. In particular, it was argued in \cite{gubser} that (for
solutions with ${\widetilde \Lambda}=0$) a 
singularity is physical only if $V$ is bounded above there.}:
\begin{equation}\label{VVV}
 V\rightarrow-\infty~{\mbox{at}}~y=y_\pm~
 {\mbox{for}}~{\widetilde \Lambda}\leq0~.
\end{equation}  
Note that this condition is necessary but might not always be sufficient
for (\ref{allowed1}) to be satisfied.
Thus, generally the full condition (\ref{allowed1}) should be used to
check whether a given singularity is physically allowed.

{}Here we note that for
${\widetilde \Lambda}>0$ the corresponding condition on $V$ is more detail
dependent. In particular, as we will see in the following, there exist 
(discontinuous) domain walls with physically acceptable singularities
and ${\widetilde\Lambda}>0$ such that $V$ is not unbounded below at 
singularities.

\begin{center}
 {\em Potentials Unbounded Above}
\end{center}

{}Let us now consider potentials unbounded above at the singularities.
Let us see if we can satisfy the condition (\ref{allowed0}).
First, the above discussion for singularities where $A\rightarrow-\infty$ 
applies to this case as well. In particular, solutions with 
${\widetilde\Lambda}\leq 0$ with singularities where $A\rightarrow -\infty$
and $V$ is unbounded above are not physically allowed.   

{}Next, let us assume
that $A$ goes to a constant at, say, $y_+$. Recall that $A^{\prime\prime}$ must
go to $-\infty$. In particular, $(\phi^\prime)^2$ goes to $\infty$ as fast
as $-A^{\prime\prime}$. Let $A^\prime\equiv N$, and $z\equiv 1/(y_+ -y)$.
As $y\rightarrow y_+$ we have $z\rightarrow+\infty$. Now, $N^\prime=-z^2 
N_z$, where the subscript $z$ denotes derivative w.r.t. $z$. 
On the other hand, $\phi^\prime=-z^2\phi_z$. This then implies that
at $z\rightarrow+\infty$ we have (see ({\ref{A''V}))
\begin{equation}
 \phi_z\sim\eta{{D-2}\over 2}{1\over z} \sqrt{N_z}~,
\end{equation}
where $\eta=\pm 1$. For $\phi$ to blow up at $z\rightarrow+\infty$ (that
is, at $y\rightarrow y_+$) it is then necessary that $N_z$ be a 
non-decreasing\footnote{More precisely, this is correct up to the
usual ``logarithmic'' factors. For instance, let $N_z\sim 1/\ln^2(z)$.
Then $\phi_z\sim 1/z\ln(z)$, and $\phi\sim\ln(\ln(z))$, so that $\phi$
blows up at $z\rightarrow +\infty$. However, such logarithmic factors
do not affect the following discussion. In particular, in this
case $N\sim z/\log^2(z)$, which blows up at large $z$.}
function of $z$. But then it follows that $N$ blows up at $z\rightarrow+\infty$
at least as fast as $z$. That is, $A^\prime$ blows up at $y\rightarrow y_+$ at
least as fast as $1/(y_+-y)$. Thus, the condition (\ref{allowed0}) 
cannot be satisfied for such domain walls
(recall that ${\cal A}^\prime=(D-1)A^\prime\exp[(D-1)A]$, so to have
finite ${\cal A}^\prime(y_+)$ we must have finite 
$A^\prime(y_+)$ as $A$ goes to a 
constant there by assumption). Similarly, if $A$ goes to a constant at $y=
y_-$, (\ref{allowed0}) cannot be satisfied there.  
We therefore conclude that singularities where $A$ goes to a constant
are unphysical (regardless of the value of ${\widetilde\Lambda}$).

\begin{center}
 {\em Physical Interpretation}
\end{center}

{}From the above discussion it follows that for singularities to be physical
they must satisfy the condition (\ref{allowed1}). Here we would like to
discuss a physical interpretation of this condition.

{}Thus, we would like to be able to cut off the space in the $y$
direction at singularities. That is, singularities correspond to boundaries
of the space. On the other hand, the condition (\ref{allowed1}) can be
rewritten as (note that ${\cal A}=\sqrt{-G}$)
\begin{equation}
 (\sqrt{-G})^\prime\Big|_{y_\pm}=0~.
\end{equation} 
That is, at a boundary the normal derivative of the volume density must
vanish. This is just as well for otherwise there would be no reason why we
should not continue the space in the $y$ direction beyond such a boundary.

{}Note that if we have a domain wall which is singular on both sides, then
topologically the space is an interval, which is the same as 
${\bf S}^1/{\bf Z}_2$.
Similarly, if the domain wall is singular on one side only, then 
topologically the space is ${\bf R}/{\bf Z}_2$. However, the actual
compactification, say, in the former case does {\em not} correspond to that on
${\bf S}^1/{\bf Z}_2$ in the sense that it cannot be viewed as an orbifold
of an ${\bf S}^1$ compactification. In particular, in this case we do not have
the usual Kaluza-Klein modes (or, more precisely, their combinations invariant
under the corresponding ${\bf Z}_2$ orbifold action) arising in the ${\bf S}^1$
compactifications.

\subsection{Discontinuous Domain Walls} 

{}In the previous subsections we discussed smooth domain walls.
Here we would like to consider discontinuous domain walls. More concretely,
both $\phi$ and $A$ could be continuous, but need not have continuous 
derivatives. In particular, to have a non-singular domain wall that localizes
gravity we must make sure that $A$ goes to $-\infty$ at $y\rightarrow\pm
\infty$. Also, in the case of singular domain walls $A$ must go to $-\infty$
at singularities. This implies that $A^\prime$ must change sign at finite $y$. 
This can happen continuously as in the previous subsections. However,
{\em a priori} we also have the following possibility\footnote{Here we should
point out that one can also consider discontinuous domain walls where 
$A^\prime$ does not change sign at the discontinuity. Furthermore, 
the following
discussion can be straightforwardly generalized to include multiple 
discontinuities.}. Let $A^\prime>0$
for $y<y_0$, and $A^\prime<0$ for $y>y_0$. At $y=y_0$ $A^\prime$ is 
discontinuous. Then $A^{\prime\prime}$ behaves as a $\delta$-function at
$y=y_0$. To be able to satisfy the equations of motion for $A$ and $\phi$ we
then must add a source term to the action $S$ in (\ref{action}):
\begin{equation}\label{source}
 S_{\rm total}=S+S_{\rm source}~,
\end{equation}
where
\begin{equation}
 S_{\rm source}=-M_P^{D-2}\int_\Sigma d^{D-1} x \sqrt{-g} f(\phi)~.
\end{equation} 
Here $\Sigma$ is the $y=y_0$ hypersurface,
\begin{equation}
 g_{\mu\nu}\equiv {\delta_\mu}^M {\delta_\nu}^N G_{MN}~,
\end{equation}
and $f(\phi)$ describes the coupling of $\phi$ to the space-time defect (often
referred to as a ``brane'') corresponding to the hypersurface $\Sigma$.
The above source term (generically) 
leads to discontinuities in $A^\prime$ and $\phi^\prime$
at $y=y_0$, and the corresponding jump conditions must be imposed on the 
solution.

{}The equations of motion now read:
\begin{eqnarray}
 && {8\over{D-2}}\nabla^2\phi- V_\phi-{\sqrt{-g}\over\sqrt{-G}}
 f_\phi\delta(y-y_0)=0~,\\
 &&R_{MN}-{1\over 2}G_{MN} R={4\over {D-2}}\left[\nabla_M\phi\nabla_N\phi
 -{1\over 2}G_{MN}(\nabla \phi)^2\right]-{1\over 2}G_{MN} V-\nonumber\\
 &&~~~~~~~~~~~~~~{1\over 2}
 {\sqrt{-g}\over\sqrt{-G}}{\delta_M}^\mu{\delta_N}^\nu g_{\mu\nu}f\delta(y-
 y_0)~.
\end{eqnarray}
As before, we are interested in solutions with the metric of the following
form:
\begin{equation}
 ds_D^2=\exp(2A){\widetilde g}_{\mu\nu} dx^\mu dx^\nu+dy^2~,
\end{equation}
where the warp factor $A$, which is a function of $y$,
is independent of the coordinates $x^\mu$,
and the $(D-1)$-dimensional metric 
${\widetilde g}_{\mu\nu}$ is independent of $y$. Moreover, $\phi$ is
independent of $x^\mu$, but can depend non-trivially on $y$. With this 
ans{\"a}tz, we have the following
equations of motion for $\phi$ and $A$:
\begin{eqnarray}\label{phi''d}
 &&{8\over {D-2}}\left[\phi^{\prime\prime}+(D-1)A^\prime\phi^\prime\right]-
 V_\phi-f_\phi\delta(y-y_0)=0~,\\
 \label{phi'A'd}
 &&(D-1)(D-2)(A^\prime)^2-{4\over{D-2}}(\phi^\prime)^2+V-
 {{D-1}\over{D-3}}{\widetilde \Lambda}\exp(-2A)=0~,\\
 \label{A''d}
 &&(D-2)A^{\prime\prime}+{4\over {D-2}}(\phi^\prime)^2+{1\over {D-3}}
 {\widetilde \Lambda}\exp(-2A)+{1\over 2} f\delta(y-y_0)=0~.
\end{eqnarray}  
The jump conditions read:
\begin{eqnarray}
 &&{8\over {D-2}}\left[\phi^\prime(y_0+)-\phi^\prime(y_0-)\right]-
 f_\phi(\phi_0)=0~,\\
 &&(D-2)\left[A^\prime(y_0+)-A^\prime(y_0-)\right]+{1\over 2}f(\phi_0)=0~,\\
 &&\phi(y_0+)=\phi(y_0-)\equiv\phi_0~,\\
 &&A(y_0+)=A(y_0-)\equiv A_0~.
\end{eqnarray}
{}From our previous discussions it should be clear that if we demand
that $A$ goes to $-\infty$ at $y=y_\pm$, then $A^\prime(y_0-)\geq 0$ and
$A^\prime(y_0+)\leq 0$, so that from the jump condition on $A^\prime$ we have
\begin{equation}
 f(\phi_0)\geq 0~.
\end{equation} 
Since $M_P^{D-2} f(\phi_0)$ is nothing but the tension of the aforementioned
space-time defect (``brane''), it then follows that gravity can be localized
in this context only if this tension is non-negative\footnote{Note that
this need not be the case if, say, $A^\prime$ 
does not change sign at the discontinuity.}.

{}It is now clear how to obtain a solution of this type 
with finite compactification volume. Start with two {\em smooth}
domain walls, call them (1) and (2), which are solutions of the equations
of motion {\em without} the source term. Let these domain walls be such that
$A^{(1)}\rightarrow-\infty$ as $y\rightarrow y^{(1)}_-$, and 
$A^{(2)}\rightarrow-\infty$ as $y\rightarrow y^{(2)}_+$. Moreover, let 
$y^{(2)}_+>y^{(1)}_-$, and assume that there is a point $y_0$, $y^{(1)}_-<y_0<
y^{(2)}_+$, such that $\phi^{(1)}(y_0)=\phi^{(2)}(y_0)\equiv\phi_0$, and 
$A^{(1)}(y_0)=A^{(2)}(y_0)\equiv A_0$ with some finite $\phi_0$ and $A_0$.
Now consider the domain wall given by
\begin{eqnarray}
 &&\phi(y)=\phi^{(1)}(y)~,~~~y^{(1)}_-<y\leq y_0~,\\
 &&\phi(y)=\phi^{(2)}(y)~,~~~y_0<y<y^{(2)}_+~,\\ 
 &&A(y)=A^{(1)}(y)~,~~~y^{(1)}_-<y\leq y_0~,\\
 &&A(y)=A^{(2)}(y)~,~~~y_0<y<y^{(2)}_+~.
\end{eqnarray}
This {\em discontinuous} domain wall will satisfy the equations of motion 
{\em with} the source term as long as the function $f(\phi)$ is such that
the following jump conditions are satisfied: 
\begin{eqnarray}
 &&{8\over {D-2}}\left[(\phi^{(2)})^\prime(y_0)-(\phi^{(1)})^\prime(y_0)
 \right]-f_\phi(\phi_0)=0~,\\
 &&(D-2)\left[(A^{(2)})^\prime(y_0)-(A^{(1)})^\prime(y_0)\right]+
 {1\over 2}f(\phi_0)=0~.
\end{eqnarray}

{}Note that from our discussion at the beginning of this section it 
follows that
discontinuous domain walls with finite ${\widetilde M}_P$ 
which are non-singular or singular on one side only
cannot have positive cosmological constant. On the other hand, 
discontinuous domain walls with finite ${\widetilde M}_P$ 
which are singular on both sides can {\em a priori} have either vanishing
or non-vanishing cosmological constant of either sign.

{}Here we can repeat the analysis of the previous subsection for discontinuous
domain walls. It is then not difficult to show that $S_{\rm total}$ on the
solution has the same form as (\ref{SSS}), so that we must impose
the consistency condition (\ref{allowed}). The latter is automatically
satisfied for non-singular domain walls, but for singular domain walls it is
non-trivial. Here we would like to give a simple example which does {\em not} 
satisfy this condition.

\begin{center}
 {\em Examples with Unphysical Singularities}
\end{center}

{}Thus, consider the case where the scalar potential vanishes:
\begin{equation}
 V(\phi)\equiv 0~.
\end{equation}
Let us look for domain wall solutions with zero cosmological constant.
Let us rewrite the scalar potential in terms of $W$:
\begin{equation}
 V=W_\phi^2-\gamma^2 W^2~.
\end{equation}
Without loss of generality we can choose 
\begin{equation}
 W=\xi\exp(\gamma\phi)~,
\end{equation}
where $\xi$ is an arbitrary constant. The smooth domain walls with ${\widetilde
\Lambda}=0$ are then the solutions to the equations
\begin{eqnarray}
 &&\phi^\prime=\alpha W_\phi=\alpha\xi\gamma\exp(\gamma\phi)~,\\
 &&A^\prime=\beta W=\beta\xi\exp(\gamma\phi)~.
\end{eqnarray}
These solutions are given by:
\begin{eqnarray}
 &&\phi=-{1\over\gamma}\ln[\alpha\xi\gamma^2(y_*-y)]~,\\
 &&A=C-{\beta\over\alpha\gamma^2}\ln[\alpha\xi\gamma^2(y_*-y)]~,
\end{eqnarray}
where $C$ is an integration constant, and $y_*$ corresponds to the location
of the singularity. As expected, at the singularity $A\rightarrow-\infty$
(recall that $\alpha$ and $\beta$ have opposite signs). Note, however, that
at the other edge $A\rightarrow +\infty$, so that these smooth domain walls
do not localize gravity.

{}Now take two such domain walls satisfying the aforementioned properties, 
and construct a discontinuous domain wall with the edges corresponding to
singularities (that is, such a domain wall is singular on both sides). This
domain wall then has finite compactification volume. However, it does not
satisfy the consistency condition (\ref{allowed}), and therefore the 
singularities are unphysical. Indeed, we have (for definiteness
we assume that $\alpha\xi>0$, so that $y<y_*$, that is, $y_*$ corresponds
to $y_+$)
\begin{equation}
 {\cal A}=\exp[(D-1)A]=\exp[(D-1)C]\alpha\xi\gamma^2(y_*-y)~,
\end{equation}  
so that 
\begin{equation}
 {\cal A}^\prime=-\exp[(D-1)C]\alpha\xi\gamma^2~.
\end{equation}
This then implies that the consistency condition (\ref{allowed0}) is not
satisfied\footnote{This observation was also made by Gary Shiu.} 
for these solutions\footnote{Here we note that such discontinuous 
domain walls were recently discussed in \cite{ASKS,KSS,erlich}.}. In fact, it
is not difficult to show that the same conclusion applies to discontinuous 
singular domain
walls with non-vanishing ${\widetilde\Lambda}$ obtained by starting 
with $V(\phi)\equiv 0$ and adding a source term\footnote{Such domain walls
were recently studied in \cite{KSS1}.}.

\subsection{Smooth {\em vs.} Discontinuous Domain Walls}

{}The purpose of this subsection is to point out that 
a discontinuous domain wall of the aforementioned type
{\em cannot} be thought of as a limit of a smooth domain wall.
The physical reason for this is actually quite simple. Thus, note that
in the case of a smooth domain wall with finite ${\widetilde M}_P$ the
localization widths for the scalar field $\phi$ and gravity are related to
each other - essentially they are both determined by the fall-off of the
warp factor $\exp(2A)$. Thus, if we take the localization width of the scalar
field to zero with the intension of reproducing the source term in 
(\ref{source}), the localization width for gravity will also go to zero. Then
for finite $D$-dimensional Planck scale $M_P$ the $(D-1)$-dimensional Planck
scale ${\widetilde M}_P$ goes to zero as well. This indicates that we cannot
approximate a discontinuous domain wall by a limit of a smooth domain wall.

{}To understand this point in a bit more detail, let us consider
the original Randall-Sundrum scenario \cite{RS}, where the scalar
potential is a negative constant
\begin{equation}
 V=-\xi^2~.
\end{equation}
Let us consider solutions with vanishing cosmological constant 
${\widetilde \Lambda}=0$. We have the following smooth solutions
($\phi_0$ is a constant):
\begin{eqnarray}
 &&\phi=\phi_0~,\\
 &&A={\beta\xi\over\gamma} y+{\rm const.}~,
\end{eqnarray}
which correspond to the $D$-dimensional AdS space-time.

{}Now let us add the source term with $f(\phi)$ independent of $\phi$:
\begin{equation}
 f(\phi)\equiv f_0~.
\end{equation}
We then find the following discontinuous domain walls with finite ${\widetilde
M}_P$ (for definiteness we are assuming $\beta\xi>0$):
\begin{eqnarray}
 &&\phi(y)=\phi_0~,\\
 &&A(y)=A_0-\beta\xi |y-y_0|~,
\end{eqnarray}
with the following jump condition:
\begin{equation}
 f_0=4(D-2)\beta\xi~,
\end{equation}
which implies that the ``brane'' tension must be fine-tuned to the
$D$-dimensional cosmological constant ($\xi=\sqrt{-V}$) for such a solution to
exist. Note that this solution is precisely the Randall-Sundrum solution
up to the extra spectator scalar field $\phi$.

{}We can now ask whether we can approximate this solution by a limit of
a smooth solution. The starting point, as it should become clear in a moment,
is not very important here as long as it correctly reproduces the features
of the above solution such as the fact that away from the ``brane'' we have
AdS vacua. Thus, for illustrative purposes, let us consider the kink
solution discussed in subsection B of section III. Now, $W$ is given
by
\begin{equation}
 W={3\xi\over 2\gamma}\left[\phi-{1\over 3}\phi^3\right]~,
\end{equation}
so that in the AdS minima with $W_\phi=0$ ($\phi=\pm 1$) we have 
$W=\pm \xi/\gamma$, and $V=-\xi^2$. The solution for $\phi$
and $A$ is given by
\begin{eqnarray}
 &&\phi(y)=\tanh\left[{3\alpha\xi\over 2\gamma}(y-y_0)\right]~,\\
 &&A={2\beta\over 3\alpha}\left[\ln\left[\cosh\left[{3\alpha\xi\over2\gamma}
 (y-y_0)\right]\right]-{1\over 4}{1\over\cosh^2\left[{3\alpha\xi\over2\gamma}
 (y-y_0)\right]}\right]+C~.
\end{eqnarray}
It is now clear that the Randall-Sundrum solution for $A$ 
is obtained in the limit
$\xi\rightarrow\infty$. (In this limit $\phi(y)=-{\rm sign}(y-y_0)$.) However,
in this limit the $D$-dimensional cosmological constant $V=-\xi^2\rightarrow
-\infty$, and the $(D-1)$-dimensional Planck scale
\begin{equation}
 {\widetilde M}_P^{D-3}={2\over{D-3}}{1\over\beta\xi} M_P^{D-2}\rightarrow 0
\end{equation}
for finite $D$-dimensional Planck scale $M_P$. Note that even if we
attempt to take $M_P$ to infinity to have ${\widetilde M}_P$ finite, the
effective $D$-dimensional cosmological constant $M_P^{D-2} V$ would still go
to $-\infty$. This indicates that discontinuous solutions cannot be 
thought of as limits of smooth solutions.

\begin{center}
 {\em Examples of Discontinuous Solutions with ${\widetilde\Lambda}\not=0$}
\end{center}

{}Before we end this section, we would like to discuss discontinuous solutions
with non-zero cosmological constant. In particular, let us consider the 
Randall-Sundrum setup. As before this is achieved by setting $\phi(y)\equiv
\phi_0$, and $V=-\xi^2$. A restricted set of solutions with 
${\widetilde \Lambda}\not=0$ in this
context were considered in \cite{nihei,kaloper,giga}. Here we would like to
discuss the most general solutions of this type\footnote{Discontinuous
domain walls with non-zero ${\widetilde\Lambda}$ were recently studied in
a more general context in \cite{odintsov}.}.

{}Thus, first consider the equations of motion for $A$ without the source
term:
\begin{eqnarray}
 &&(D-1)(D-2) (A^\prime)^2=\xi^2 +{{D-1}\over {D-3}}{\widetilde \Lambda}
 \exp(-2A)~,\\
 &&(D-2)A^{\prime\prime}=-{1\over{D-3}}{\widetilde\Lambda}\exp(-2A)~.
\end{eqnarray}
Let
\begin{equation}
 \exp(A)\equiv U~.
\end{equation}
Then we have:
\begin{eqnarray}
 &&(D-1)(D-2) (U^\prime)^2=\xi^2 U^2 +{{D-1}\over{D-3}}{\widetilde \Lambda}~,\\
 &&(D-1)(D-2) U^{\prime\prime}=\xi^2 U~.
\end{eqnarray}
The general solution to this system of equations is given by
\begin{equation}
 U=C\left[\exp(ay)-{{\widetilde\Lambda}\over 4\xi^2 C^2}\exp(-ay)\right]~,
\end{equation}
where 
\begin{equation}
 a\equiv{\xi\over\sqrt{(D-1)(D-2)}}~,
\end{equation}
and $C$ is an integration constant.

{}Note that if ${\widetilde\Lambda}<0$, then we must choose $C>0$, and
the solution is non-singular. Moreover, $U$ blows up at both $y\rightarrow\pm
\infty$, so even if we add the source term, the corresponding solution
cannot localize gravity. On the other hand, for ${\widetilde\Lambda}>0$ $U$
vanishes for some finite $y$, so that the domain wall is singular. Note that
without loss of generality we can assume that $C>0$. Then we can obtain a
singular domain wall with non-vanishing ${\widetilde \Lambda}$ and finite
${\widetilde M}_P$ by adding the source term and gluing together the 
appropriate parts of two smooth domain walls. Thus, consider the solution
(for definiteness we assume $a>0$, that is, $\xi>0$)
\begin{equation}\label{U}
 U(y)=C\left[\exp(-a|y|)-{{\widetilde \Lambda}\over4\xi^2 C^2}\exp(a|y|)
 \right]~,
\end{equation}
where
\begin{equation}
 C>{\sqrt{\widetilde \Lambda}\over2\xi}~.
\end{equation}
The singularities (which are physical as they satisfy (\ref{allowed1}))
are located at
\begin{equation}
 y_\pm=\pm{1\over a}\ln\left[{2\xi C\over\sqrt{\widetilde\Lambda}}\right]~.
\end{equation}
The discontinuity is located at $y_0=0$, and the corresponding jump condition 
reads:
\begin{equation}
 f_0=4(D-2)a\coth(a|y_\pm|)>0~.
\end{equation}
Here we note that the solutions discussed in \cite{nihei,kaloper,giga} 
correspond to taking $C-{\widetilde\Lambda}/4\xi^2 C=1$. However, generally
$C$ is an independent integration constant. In particular, {\em a priori}
${\widetilde\Lambda}$ is not determined by specifying $\xi$ and
$f_0$. However, the ratio of the effective $(D-1)$-dimensional cosmological 
constant ${\widetilde M}_P^{D-3} {\widetilde\Lambda}$ to 
${\widetilde M}_P^{D-1}$, that is, ${\widetilde \Lambda}/{\widetilde M}_P^2$,
depends on ${\widetilde \Lambda}$ and $C$ via the ratio 
${\widetilde\Lambda}/C^2$, which is fixed via $\xi$ and $f_0$.  

{}Next, consider the same $U(y)$ as in (\ref{U}) with 
\begin{equation}
 C<{\sqrt{\widetilde \Lambda}\over2\xi}~.
\end{equation}
It then describes a non-singular domain wall which does {\em not} localize
gravity. In fact, in this case the jump condition reads:
\begin{equation}
 f_0=-4(D-2)a\coth(\zeta)<0~,
\end{equation}
where 
\begin{equation}
 \zeta\equiv{1\over a}\ln\left[{\sqrt{\widetilde\Lambda}\over 2\xi C}\right]~.
\end{equation}
Note that such a domain wall would require a negative tension ``brane''.

\subsection{Summary}

{}In this subsection we would like to briefly summarize the results
of this section. 

{}(i) Non-singular smooth domain walls with $A\rightarrow-\infty$ 
at $y\rightarrow\pm\infty$
(that is,
finite compactification volume and ${\widetilde M}_P$) cannot have 
${\widetilde \Lambda}>0$. On the other hand, non-singular smooth domain
walls approaching AdS vacua on at least one side can only have ${\widetilde
\Lambda}=0$.

{}(ii) Singular smooth domain walls can {\em a priori} have vanishing
or non-vanishing ${\widetilde \Lambda}$ of either sign.
Singularities with $A\rightarrow-\infty$ are physical only if
the scalar potential is unbounded below there. However, singularities where 
$A\rightarrow{\rm const.}$, which can only 
arise if the potential is unbounded above there, are unphysical. In particular,
it is not consistent to cut off the space at such singularities in the $y$
direction.

{}(iv) The aforementioned conclusions also apply to discontinuous domain 
walls (with a $\delta$-function source term). However, such a domain wall
cannot be thought of as a limit of a smooth domain wall with non-zero 
${\widetilde M}_P$.

\section{Normalizable Modes}

{}In this section we would like to discuss normalizable modes living 
on the domain walls discussed in the previous sections. In particular, let us
focus on non-singular domain walls. To study normalizable modes, it is 
convenient to consider small fluctuations around the background metric 
$G_{MN}$ and the scalar field $\phi$, which we will denote by $h_{MN}$ and 
$\varphi$, respectively. Using the diffeomorphism invariance of the theory we
can choose the following gauge:
\begin{equation}\label{gauge}
 h_{\mu D}=h_{DD}=0~.
\end{equation}
Instead of the remaining metric fluctuations $h_{\mu\nu}$, it will be 
convenient to work with ${\widetilde h}_{\mu\nu}$ defined via
\begin{equation}
 h_{\mu\nu}=\exp(2A) {\widetilde h}_{\mu\nu}~.
\end{equation}
The quadratic action for ${\widetilde h}_{\mu\nu}$ and $\varphi$ is given by
\begin{eqnarray}
 &&S[{\widetilde h}_{\mu\nu},\varphi]/M_P^{D-2}=\nonumber\\
 &&\int d^D x
 \exp[(D-3)A] \left\{ \left(\sqrt{-{\widehat g}}
 \left[{\widehat R}-{\widetilde \Lambda}\right]\right)^{(2)}-
 \sqrt{-{\widetilde g}}{1\over\alpha^2}
 ({\widetilde \nabla}\varphi)^2\right\}-\nonumber\\
 &&\int d^D x
 \exp[(D-1)A] \sqrt{-{\widetilde g}}\left\{{1\over\alpha^2}(\varphi^\prime)^2
 +{1\over 2}V_{\phi\phi} \varphi^2 -{1\over\alpha^2}\phi^\prime
 {\widetilde h}^\prime\varphi +{1\over 4}\left[
 ({\widetilde h}^\prime_{\mu\nu})^2
 -({\widetilde h}^\prime)^2\right]\right\}~.
\end{eqnarray}
Here we are using the following notations. First, ${\widehat g}\equiv
\det({\widehat g}_{\mu\nu})$, where ${\widehat g}_{\mu\nu}\equiv {\widetilde
g}_{\mu\nu}+{\widetilde h}_{\mu\nu}$. Also, ${\widehat R}$ is the 
$(D-1)$-dimensional Ricci
scalar constructed from the metric ${\widehat g}_{\mu\nu}$. The superscript 
``$(2)$'' in the term 
\begin{equation}
 \left(\sqrt{-{\widehat g}}
 \left[{\widehat R}-{\widetilde \Lambda}\right]\right)^{(2)}
\end{equation}
indicates that only the terms quadratic in ${\widetilde h}_{\mu\nu}$ should
be kept in the corresponding expression. Finally, ${\widetilde h}\equiv
{\widetilde h}^\mu_\mu$, and $({\widetilde h}^\prime_{\mu\nu})^2\equiv
({\widetilde h}_{\mu\nu})^\prime({\widetilde h}^{\mu\nu})^\prime$, where the
indices are lowered and raised with the metric ${\widetilde g}_{\mu\nu}$ and
its inverse. We also note that the boundary contributions discussed in 
subsection C of section V vanish in the case of non-singular domain walls.

{}As in the case of non-gravitational domain walls, the above action 
has a much simpler form if instead of $\varphi$ we use the field $\omega$
defined via
\begin{equation}
 \varphi\equiv\phi^\prime\omega~.
\end{equation}
Thus, the quadratic action for ${\widetilde h}_{\mu\nu}$ and $\omega$ is
given by:
\begin{eqnarray}\label{homega}
 &&S[{\widetilde h}_{\mu\nu},\omega]/M_P^{D-2}=\nonumber\\
 &&\int d^D x
 \exp[(D-3)A] \left\{ \left(\sqrt{-{\widehat g}}
 \left[{\widehat R}-{\widetilde \Lambda}\right]\right)^{(2)}-
 \sqrt{-{\widetilde g}}{1\over\alpha^2}(\phi^\prime)^2 \left[
 ({\widetilde \nabla}\omega)^2-{{D-1}\over{D-2}}{{\widetilde \Lambda}\over
 {D-3}}\omega^2\right]\right\}-\nonumber\\
 &&\int d^D x
 \exp[(D-1)A] \sqrt{-{\widetilde g}}\left\{{1\over\alpha^2}(\phi^\prime)^2
 (\omega^\prime)^2
 +{1\over 4}\left[
 ({\widetilde h}^{*\prime}_{\mu\nu})^2
 -{{D-2}\over{D-1}}\left({\widetilde h}^\prime+2\gamma^2(\phi^\prime)^2
 \omega\right)^2\right]\right\}~.
\end{eqnarray}
Here 
\begin{equation}
 {\widetilde h}^*_{\mu\nu}\equiv{\widetilde h}_{\mu\nu}-{1\over{D-1}}
 {\widetilde h} {\widetilde g}_{\mu\nu}
\end{equation}
is the traceless part of ${\widetilde h}_{\mu\nu}$.

{}Next, we would like to study normalizable modes. Here we are assuming that
the warp factor $\exp(2A)$ goes to zero at $y\rightarrow\pm\infty$ fast enough,
so that the compactifications volume as well as ${\widetilde M}_P$ are finite.
Then normalizable modes correspond to configurations with $\omega^\prime=0$.
It is not difficult to show that 
the equation of motion for such $\omega$ simplifies as follows:
\begin{equation}\label{h'}
 {\widetilde h}^\prime=-2\exp(-2A){\widetilde\nabla}^2\omega+2(D-1)
 A^{\prime\prime} \omega~.
\end{equation}
Note that this equation does not explicitly contain ${\widetilde\Lambda}$.

{}Thus, the equation of motion (\ref{h'}) for $\omega=\omega(x^\mu)$
seems to imply that we have normalizable modes satisfying the 
$(D-1)$-dimensional Klein-Gordon equation 
\begin{equation}
 {\widetilde\nabla}^2\omega=m^2\omega
\end{equation} 
for any $m^2$ (including massive and tachyonic). At first this might
appear to imply that the domain wall is unstable. This is, however, not the
case due to the following. Note that $\omega$ (or, equivalently, $\varphi$)
mixes with ${\widetilde h}^\prime$, that is, the $y$ derivative of the trace 
part of ${\widetilde h}_{\mu\nu}$. 
Moreover, as we will see in a moment, there is a residual
diffeomorphism invariance which respects the gauge choice (\ref{gauge}). 
These facts have important implications which we would like to discuss next.

\subsection{Residual Diffeomorphism Invariance}

{}The gauge choice (\ref{gauge}) does not fix all the gauge freedom in the
system. Thus, there is a residual diffeomorphism invariance which preserves
(\ref{gauge})\footnote{This residual gauge invariance was discussed in the
context of the Randall-Sundrum model in \cite{GT}, and also in 
\cite{giddings}.}. 
For our purposes here it will suffice to consider infinitesimal
residual diffeomorphisms. Thus, we have
\begin{equation}
 \delta h_{MN}=\nabla_M\xi_N+\nabla_N\xi_M~.
\end{equation}
In particular,
\begin{equation}
 \delta h_{DD}=2\xi_D^\prime~,
\end{equation}
and the gauge condition $h_{DD}=0$ is preserved if and only if
\begin{equation}
 \xi_D^\prime=0~.
\end{equation}
Furthermore, we have
\begin{equation}
 \delta h_{\mu D}={\widetilde \nabla}_\mu \xi_D +\xi_\mu^\prime-2A^\prime
 \xi_\mu~,
\end{equation}
where on the r.h.s. we have kept only the terms independent of $h_{MN}$, which
is consistent with the linearized approximation. Requiring that the gauge
condition $h_{\mu D}=0$ is preserved, we obtain
\begin{equation}
 \xi_\mu=\exp(2A){\widetilde \xi}_\mu~,
\end{equation}
where ${\widetilde \xi}_\mu$ satisfies the following first order differential
equation
\begin{equation}
 {\widetilde\xi}_\mu^\prime=-\exp(-2A) {\widetilde\nabla}_\mu\xi_D~.
\end{equation}
Finally, for the rest of the $h_{MN}$ components we obtain
\begin{equation}
 \delta h_{\mu\nu}={\widetilde\nabla}_\mu\xi_\nu+
 {\widetilde\nabla}_\nu\xi_\mu +2A^\prime\exp(2A)
 \xi_D {\widetilde g}_{\mu\nu}~,
\end{equation}
or, equivalently,
\begin{equation}
  \delta {\widetilde h}_{\mu\nu}={\widetilde\nabla}_\mu{\widetilde \xi}_\nu+
 {\widetilde\nabla}_\nu{\widetilde \xi}_\mu +2A^\prime
 \xi_D {\widetilde g}_{\mu\nu}~.
\end{equation}
In particular, we have
\begin{equation}
 \delta{\widetilde h} =2{\widetilde\nabla}^\mu{\widetilde\xi}_\mu +2(D-1)
 A^{\prime} \xi_D~,
\end{equation}
from which it follows that
\begin{equation}
 \delta{\widetilde h}^\prime=-2\exp(-2A){\widetilde\nabla}^2\xi_D+
 2(D-1)A^{\prime\prime}\xi_D~.
\end{equation}

{}The last result implies that for any choice of $\omega$ in (\ref{h'})
(where $\omega$ is assumed to be independent of $y$), the corresponding
${\widetilde h}$ is a pure gauge, and the corresponding gauge parameter
is given by
\begin{equation}
 \xi_D=\omega~.
\end{equation}
That is, $\omega$ is {\em not} a propagating degree of freedom, but a gauge
parameter, and, up to a gauge transformation, ${\widetilde h}$ is independent 
of $y$. This result has a simple physical interpretation to which we now turn.

\subsection{Localization of Gravity as a Higgs Mechanism}

{}The above analyses suggest that localization of gravity is actually a
Higgs mechanism for the graviton. In particular, in the $D$-dimensional
language we have the fields $h_{MN}$ and $\varphi$ with $D(D-3)/2$ and 1
physical degrees of freedom, respectively. In the $(D-1)$-dimensional language
together with the scalar $\varphi$ we have the fields
$h_{\mu\nu}$, $h_{\mu D}$ and $h_{DD}$ corresponding to the $(D-1)$-dimensional
graviton, a graviphoton and a scalar, respectively. If we turn off the coupling
to gravity, there is a $\varphi$ zero mode 
(which might or might not be normalizable in this context)
corresponding to the Goldstone mode
of the broken translational invariance in the $y$ direction. However, once
we gauge diffeomorphisms, that is, once we consider the coupling to gravity,
the $\varphi$ zero mode is eaten by the graviphoton, which now becomes massive.
The scalar $h_{DD}$ is also massive - the volume of the compactification,
which is finite, is not a free parameter but is completely fixed in terms
of the parameters in the scalar potential $V$. This is analogous to what 
happens, say, in the Abelian Higgs model, where a $U(1)$ gauge field is coupled
to a single complex scalar field. Once the latter acquires an expectation 
value, the angular part is eaten by the gauge field, which therefore becomes
massive, while the radial part acquires a mass due to a non-trivial scalar
potential. However, as we will see in a moment, the Higgs mechanism that 
takes place in the localization of gravity has certain novel features, which
is due to important differences between gravity and an ordinary gauge 
theory.

\subsection{Zero and Massive Modes} 

{}To analyze normalizable modes, consider the standard factorized form of the
corresponding field, which is ${\widetilde h}_{\mu\nu}$ in this case:
\begin{equation}
 {\widetilde h}_{\mu\nu} (x^\lambda,y)=\zeta_{\mu\nu}(x^\lambda) \Sigma(y)~.
\end{equation}
Using the fact that $\omega$ in ({\ref{h'}) is a gauge parameter, without
loss of generality we can set it to zero, which implies 
${\widetilde h}^\prime=0$, and, consequently, either $\Sigma^\prime=0$, or
$\zeta^\mu_\mu=0$. In the former case we have a zero mode
\begin{equation}
 {\widetilde h}_{\mu\nu} (x^\lambda,y)={\rm const.}\times
 \zeta_{\mu\nu}(x^\lambda)~,
\end{equation}
where $\zeta_{\mu\nu}$ satisfies the equation of motion for the massless
$(D-1)$-dimensional graviton, which is nothing but the Euler-Lagrange
equation obtained by varying the quadratic $(D-1)$-dimensional action
\begin{equation}
 {\widetilde S}[{\widetilde h}_{\mu\nu}]\equiv{\widetilde M}_P^{D-3}
 \int d^{D-1}x \left(\sqrt{-{\widehat g}}
 \left[{\widehat R}-{\widetilde \Lambda}\right]\right)^{(2)}
\end{equation}
w.r.t. ${\widetilde h}_{\mu\nu}$ (or, equivalently, $\zeta_{\mu\nu}$).
In this case the zero mode is quadratically normalizable, that is, the
$(D-1)$-dimensional Planck mass ${\widetilde M}_P$ is finite.

{}Next, if $\Sigma^\prime\not=0$, then using ({\ref{homega}) it is
not difficult to show that 
$\Sigma(y)$ must satisfy the equation 
\begin{equation}\label{Sigma}
 \left(\exp[(D-1)A]\Sigma^\prime\right)^\prime+m^2\exp[(D-3)A]\Sigma=0~,
\end{equation}
where $m^2$ is a constant. As to $\zeta_{\mu\nu}$, it then 
satisfies the equation of motion of a $(D-1)$-dimensional 
symmetric {\em traceless} (recall that ${\widetilde h}=0$ in this case) 
two-index tensor field with mass squared $m^2\not=0$ 
propagating in the background metric ${\widetilde g}_{\mu\nu}$. 
Using (\ref{Sigma}) the action for ${\widetilde h}_{\mu\nu}$ can be written as
\begin{equation}
 {\widetilde S}[{\widetilde h}_{\mu\nu}]\equiv M_P^{D-2}
 \int d^{D-1}x dy \exp[(D-3)A] \left\{\left(\sqrt{-{\widehat g}}
 \left[{\widehat R}-{\widetilde \Lambda}\right]\right)^{(2)}-{1\over 4}
 m^2 ({\widetilde h}^*_{\mu\nu})^2\right\}~,
\end{equation}
and the norm of ${\widetilde h}_{\mu\nu}$ is proportional to
\begin{equation}\label{normhhat}
 ||{\widetilde h}_{\mu\nu}||^2 \sim \int dy \exp[(D-3)A]\Sigma^2~.
\end{equation} 
We would now like to show that massive modes with $m^2>0$ are 
plane-wave normalizable (in contrast to the zero mode, which is quadratically
normalizable), while tachyonic modes with $m^2<0$ are not normalizable.

{}To see this, let us make the coordinate transformation $y\rightarrow z$
so that the metric takes the form:
\begin{equation}
 ds^2_D=\exp(2A)\left({\widetilde g}_{\mu\nu}dx^\mu dx^\nu+dz^2\right)~.
\end{equation}
That is, 
\begin{equation}
 dy=\exp(A)dz~, 
\end{equation}
where we have chosen the overall sign so that $z\rightarrow\pm\infty$ as
$y\rightarrow\pm\infty$. In this new coordinate system (\ref{Sigma}) reads:\
\begin{equation}
 \Sigma_{zz}+(D-2)A_z\Sigma_z+m^2\Sigma=0~,
\end{equation}
where the subscript $z$ denotes derivative w.r.t. $z$. Let
\begin{equation}
 \Sigma\equiv \exp\left[-{1\over 2}(D-2) A\right]{\widehat \Sigma}~.
\end{equation}
The equation for ${\widehat\Sigma}$ reads:
\begin{equation}\label{Sigmahat}
 {\widehat\Sigma}_{zz}+\left[m^2 -{1\over 2}(D-2) A_{zz}-{1\over 4}
 (D-2)^2 (A_z)^2\right]{\widehat\Sigma}=0~.
\end{equation}
Let us assume that $m^2\not=0$. Then for large $z$ the $m^2$ term is dominant
in the square brackets in (\ref{Sigmahat}) for
\begin{eqnarray}
 &&A_z=A^\prime\exp(A)\rightarrow 0~,\\
 &&A_{zz}=\left[A^{\prime\prime}+(A^\prime)^2\right]\exp(2A)\rightarrow 0
\end{eqnarray}
for large $y$ (that is, large $z$). This then implies that for $m^2>0$
${\widehat\Sigma}$ has the following asymptotic behavior at large $z$:
\begin{equation}
 {\widehat \Sigma}(z)\sim C_1\sin(mz)+C_2\cos(mz)~,
\end{equation}
where $C_1,C_2$ are some constants. On the other hand, in terms of ${\widehat
\Sigma}$ the norm of ${\widetilde h}_{\mu\nu}$ (\ref{normhhat}) 
is given by
\begin{equation}
 \int dy \exp[(D-3)A]\Sigma^2=\int dz {\widehat \Sigma}^2~.
\end{equation}
This implies that massive ${\widetilde h}_{\mu\nu}$ modes are plane-wave
normalizable (but not quadratically normalizable). Thus, we have a
continuum of massive bulk modes just as in the Randall-Sundrum model.

{}What about the tachyonic modes? It is not difficult to see that these
are not normalizable. Indeed, (\ref{Sigmahat}) can be written as 
\cite{Freedman}
\begin{equation}\label{QQ}
 (QQ^\dagger) {\widehat\Sigma}=m^2{\widehat \Sigma}~,
\end{equation}
where 
\begin{eqnarray}
 &&Q\equiv\partial_z +{1\over 2}(D-2) A_z~,\\
 &&Q^\dagger =-\partial_z+{1\over 2}(D-2) A_z~.
\end{eqnarray}
Evidently (\ref{QQ}) does not have normalizable solutions with $m^2<0$
as it is nothing but Schr{\"o}dinger's equation of a supersymmetric quantum
mechanics with the Hamiltonian ${\cal H}\equiv QQ^\dagger$ \cite{Freedman}.

{}Before we end this section, let us make the following remark. The existence
of the quadratically normalizable zero mode as well as plane-wave normalizable
massive bulk modes in the above setup is similar to what happens in the 
original Randall-Sundrum model. However, localization of gravity in the case
of smooth domain walls arises via a spontaneous breakdown of diffeomorphism
invariance, while in the Randall-Sundrum model as well as the cases where we
have discontinuous domain walls diffeomorphism invariance is explicitly broken
by the ``brane''. That is, in the former case we have Higgs mechanism, while 
in the latter case we do not, and, in particular, there is no massless
graviphoton to begin with. This is precisely the underlying reason why 
discontinuous domain walls cannot be thought of as limits of smooth domain 
walls.

\section{The Upshot}

{}In the previous sections we have studied possible types of domain wall
solutions of (\ref{action}) which localize gravity. In this section we
would like to discuss some aspects of the cosmological constant problem within
the context of a brane world realized via such a domain wall. In particular,
for the reasons which should be clear from our previous discussions, we
will focus on smooth domain walls.

{}To begin with, let us consider non-singular domain walls with infinite 
tension. In particular, suppose the corresponding scalar potential is such 
that it does not have local minima and is unbounded below.
Then it is clear that {\em no} solution that does not localize gravity is
consistent. However, as we saw in the previous sections, subject to 
non-singularity conditions on the scalar potential, there are (classically)
consistent domain wall solutions which localize gravity. This implies that 
the theory will ultimately\footnote{Here we should
point out that {\em a priori} 
there might also exist consistent localized solutions of codimension
$r>1$ if we have $r$ scalar fields. In such a case 
the corresponding theory might also be able to choose such backgrounds.} end up
choosing such a background. In such cases we therefore have {\em spontaneous}
localization of gravity via the Higgs mechanism discussed in subsection B of 
the previous section.

{}Had it been the case that domain wall solutions with infinite tension could
only have vanishing $(D-1)$-dimensional cosmological constant, the 
aforementioned property {\em a priori} might have been  
considered as a possible solution to the cosmological
constant problem. However, as we saw in the previous sections, infinite
tension domain walls can have negative cosmological constant. Could we then 
conclude that such domain walls solve ``half'' of the cosmological constant
problem as they do not allow ${\widetilde \Lambda}>0$?

{}Here we would like to point out that such a conclusion might be premature.
The reason is twofold. First, to have a non-singular domain wall, the 
scalar potential must satisfy certain conditions discussed in the previous
sections. Quantum corrections can {\em a priori} modify the potential in such a
way that these conditions are no longer satisfied. For singular domain walls,
on the other hand, non-zero cosmological constant of either sign is {\em a
priori} allowed. Nonetheless, one could in principle argue that (at least in
some sense) the subset of potentials that satisfy non-singularity conditions  
has non-zero measure, so this particular objection might be avoidable.

{}There is, however, another point here, which could potentially be more
problematic. Quantum corrections modify not only the scalar potential, but
generically also introduce higher derivative, in particular, higher
curvature terms, which become important in such backgrounds. Thus, for 
instance, the $D$-dimensional Ricci scalar is given by:
\begin{equation}
 R={\widetilde R}\exp(-2A)-(D-1)\left[2A^{\prime\prime}+D(A^\prime)^2\right]~,
\end{equation}
where ${\widetilde R}$ is the $(D-1)$-dimensional Ricci scalar. Even if
${\widetilde R}=0$ on the solution (that is, if the $(D-1)$-dimensional
cosmological constant is vanishing), the $D$-dimensional Ricci scalar blows
up at large $y$ unless $A^\prime$ goes to a constant there\footnote{Note that
if on the solution
${\widetilde R}\not=0$, then $R$ ultimately blows up at large $y$ as long as
$A$ goes to $-\infty$ there.}. In the case
of infinite tension domain walls $A^\prime$ diverges at large $y$, and so does
the Ricci scalar $R$, so that higher curvature (or, more generally, higher 
derivative) terms cannot be ignored. What is worse, in this case it is
not even clear if there exists a controlled approximation scheme for taking 
into account such terms. The underlying reason for this is that such domain
walls are actually singular but the singularities are located at $y\rightarrow
\pm \infty$ (that is, at infinite distances from the ``core'' of the domain 
wall). In particular, {\em a priori} there does not seem to be a reason to
believe that any conclusions about allowed values of the $(D-1)$-dimensional
cosmological constant would not be modified by higher curvature terms. In fact,
as we will point out in the following, in this context it is not even 
clear whether such domain walls do indeed localize gravity. Note that the
above discussion also applies to singular domain walls with singularities
located at finite values of $y$. 

{}The above discussion suggests that we consider solutions where the 
$D$-dimensional Ricci scalar $R$ goes to constant values 
at large $y$. Such solutions
should have vanishing $(D-1)$-dimensional cosmological constant, and $A^\prime$
should go to constant values at large $y$. In particular, if the cut-off scale
for higher derivative terms in the action is of order $M_*$, then for such 
solutions we have a controlled approximation scheme for including higher 
curvature terms as long as 
\begin{equation}
 |R|\sim (A^\prime)^2 \ll M_*^2~.
\end{equation}
In fact, from our previous discussions it should be clear that such solutions
must necessarily interpolate between two adjacent local AdS 
minima of the scalar
potential. However, if such minima are located at finite values of $\phi$, then
{\em a priori} there exist other solutions with non-zero cosmological 
constant\footnote{If the scalar potential is bounded below, then before taking 
into account higher derivative terms such solutions are
necessarily singular and have positive cosmological constant. Otherwise one can
{\em a priori} have singular or non-singular solutions with negative 
cosmological constant. However, we should mention that 
these conclusions might be 
modified by higher curvature terms.}. These solutions have the property that
$\phi$ ``overshoots'' the values corresponding to the local AdS minima. To 
avoid this, one could consider potentials which have no local minima but
approach constant negative values at large $\phi$.

\subsection{Runaway Potentials} 

{}Thus, we are lead to consider the following setup. Let the scalar potential,
such that it has no local minima, be of the runaway type, that is, it 
approaches constant values at large $\phi$. More precisely, let these values
be negative, and satisfy the condition
\begin{equation}
 |V(\phi\rightarrow\pm\infty)|\ll M_*^2~,
\end{equation}
with $M_*$ defined as above. Then the corresponding domain walls necessarily
have vanishing cosmological constant. In fact, it is not difficult to see that
this statement holds even if we take into account higher curvature (or, more
generally, higher derivative) terms as these are now under 
control\footnote{More precisely, this is correct subject to the convergence of
the $|R|/M_*^2$ expansion.}. We therefore conclude that for such domain walls
(at least in some sense - see below) the higher curvature corrections do not
change the conclusion that the cosmological constant must vanish. There is,
however, the following issue here. {\em A priori} there does not seem to be a
reason to believe that quantum corrections will not modify the behavior of the
scalar potential at large $\phi$. In particular, if instead of constant 
values the potential goes to $+\infty$ or $-\infty$ at large $\phi$, then
the conclusion that the cosmological constant must vanish would be 
modified\footnote{Actually, if the potential is of the aforementioned 
runaway type on at least one side, it is enough to ensure vanishing of the
cosmological constant.}. Note that supersymmetry locally broken in the ``core''
of the domain wall does not 
seem to help here - since the compactification volume is actually finite,
supersymmetry breaking in the bulk is non-vanishing as it is only
suppressed by the volume factor.

{}At any rate, if the runaway behavior of the scalar potential is not lifted
by quantum corrections\footnote{Here we should mention that typically it is 
not easy to lift runaway behavior perturbatively.}, then such domain
walls could {\em a priori} lead to a solution to the cosmological constant 
problem. 

{}Here we would like to discuss an explicit example of such a domain
wall. Since the domain wall has vanishing ${\widetilde \Lambda}$, it is
convenient to express the scalar potential in terms of $W$ via (\ref{super}).
Thus, consider the example where
\begin{equation}
 W=\xi\tanh(\phi)~.
\end{equation}
The scalar potential reads:
\begin{equation}
 V=\xi^2\left[{1\over\cosh^4(\phi)}-\gamma^2\tanh^2(\phi)\right]~,
\end{equation}
and it approaches a constant value $V\rightarrow-\gamma^2\xi^2$ at $\phi
\rightarrow\pm\infty$.
The solution for $\phi$ and $A$ is given by
\begin{eqnarray}
 &&2\phi+\sinh(2\phi)=4\alpha\xi (y-y_0)~,\\
 &&A=C+{\beta\over 4\alpha}\cosh(2\phi)~,
\end{eqnarray}
where $y_0$ and $C$ are integration constants. Note that $|\phi|$ grows
logarithmically for large $|y|$, whereas $A$ linearly goes to $-\infty$. In 
particular, higher derivative terms are either vanishing at large $y$, or 
subleading provided that $\xi^2\ll M_*^2$.

\subsection{Delocalization of Gravity}

{}The discussion in the previous subsection might appear promising in the
context of the cosmological context problem. There is, however, an additional
issue here, which we would like to discuss next.

{}The point is that higher curvature terms appear to delocalize gravity. 
Indeed, let us, say, consider adding to (\ref{action}) 
a generic term of the form
\begin{equation}
 \zeta \int d^D x\sqrt{-G} R^k~,
\end{equation}
where $\zeta$ has dimension of $({\rm mass})^{D-2k}$. We have
\begin{equation}
 R^k={\widetilde R}^k\exp(-2kA)+\dots~,
\end{equation} 
where the ellipses stand for terms containing lower powers of ${\widetilde
R}$. The contribution of the ${\widetilde R}^k$ term to the action then reads:
\begin{equation}
 \zeta\int d^{D-1}x dy \exp[(D-1-2k)A]\sqrt{-{\widetilde g}}{\widetilde R}^k~.
\end{equation}
Assuming that $A$ goes to $-\infty$ at large $y$, for large enough $k$ the
corresponding $y$ integral
\begin{equation}
 \int dy \exp[(D-1-2k)A]
\end{equation}
diverges, which implies that gravity is no longer localized if we include such
a higher curvature term. In fact, for $D=5$ delocalization of gravity 
generically occurs already at the four-derivative level, that is, once
we include $R^2$, $R_{MN}R^{MN}$ and $R_{MNST}R^{MNST}$ terms. In particular,
it is not difficult to show that once we include higher curvature terms there
is no normalizable graviton zero mode, and we no longer have 
$(D-1)$-dimensional
gravity localized on the wall, in particular, $(D-1)$-dimensional Newton's 
law is no longer reproduced\footnote{Here we could ask how is gravity modified.
However, generically inclusion of four-derivative terms leads to 
negative-norm 
states in the Hilbert space of the theory thus violating unitarity.}.

{}As we have already mentioned, in the case of runaway potentials considered 
in the previous subsection we have control over higher derivative terms as far
as the background is concerned. However, what we see here is that we do not
have control over the fluctuations around such a background. The reason for 
this is that at large $y$ we approach the horizon of the AdS space where 
gravity is strongly coupled. It is unclear whether there is any 
``non-perturbative'' approach to this problem, but in any case the requirement
that one reproduce the $(D-1)$-dimensional Einstein gravity at low energies 
seems reasonable\footnote{Here we would like to point out a possible way out
of this difficulty. If the warp factor $A$ goes to $+\infty$ instead of 
$-\infty$ at large $y$, then in, say, $D=5$ terms with more than four
derivatives are now normalizable. On the other hand, one then must arrange
for the lower-derivative terms to be normalizable as well, which 
{\em a priori} does not
seem impossible. Such a framework is interesting as here the volume of the
fifth dimension is actually infinite. As was pointed out in 
\cite{DGP,EW,DVALI},
in the supersymmetric context this could have important implications for the
cosmological constant problem. In particular, supersymmetry locally broken 
on a ``brane'' ({\em e.g.}, the ``core'' of a smooth domain wall) is not
transmitted to the bulk as supersymmetry breaking in the bulk is
suppressed by the compactification volume, which is infinite. For a recent
discussion of infinite volume extra dimension scenarios, see 
\cite{GRS,CEH,DGP,DGP1}.}.  

{}Before we end this subsection, we would like to make the following remark. 
If we forget about higher curvature terms beyond four derivatives, then
for $D=5$ there is an exception to the aforementioned observation that gravity
is delocalized. In particular, consider adding the Gauss-Bonnet term to the
5-dimensional action
\begin{equation}
 \zeta\int d^5 x \sqrt{-G}\left[R^2-4R_{MN}R^{MN}+R_{MNST}R^{MNST}\right]~.
\end{equation}
The corresponding term containing four derivatives w.r.t. the 4-dimensional
coordinates $x^\mu$ then reads:
\begin{equation}
 \zeta\int d^4 x dy \sqrt{-{\widetilde g}}\left[{\widetilde R}^2-4
 {\widetilde R}_{\mu\nu}{\widetilde R}^{\mu\nu}+{\widetilde R}_{\mu\nu
 \sigma\tau}{\widetilde R}^{\mu\nu\sigma\tau}\right]~.
\end{equation}
Note that the integral over $y$ is divergent - there is no warp factor
in this expression. However, this term is harmless as long as the 
four-dimensional space-time is topologically trivial. Indeed, the Gauss-Bonnet
term in four dimensions is a total derivative.

{}Note that if we considered adding the Gauss-Bonnet term in $D>5$, then 
the analog of the aforementioned term would be normalizable as the $y$ 
integral is now non-trivially weighted:
\begin{equation}
 \int dy \exp[(D-5)A]~.
\end{equation} 
Note, however, that if we expand around the $(D-1)$-dimensional Minkowski 
space,
the term quadratic in ${\widetilde h}_{\mu\nu}$ is a total derivative 
\cite{Zwiebach,Zumino} (albeit the full Gauss-Bonnet term in $D-1>4$ dimensions
no longer is). This implies that the $(D-1)$-dimensional 
graviton zero mode is unaffected even if $D>5$.

{}Discontinuous domain wall solutions in the presence of the Gauss-Bonnet term 
were recently studied in \cite{Zee}. Smooth domain walls in this context
appear to have some interesting properties which will be discussed elsewhere
\cite{olindo}\footnote{Here we should point out the following issue arising
in this context. As we have already mentioned, if we expand the
($D$-dimensional) Gauss-Bonnet
term around a flat background, the term quadratic in $h_{MN}$ is a total
derivative. In particular, it does not lead to violation of unitarity unlike
a generic four-derivative action. However, in a non-trivial warped
background the linearized theory has a non-trivial propagator. Nonetheless,
we expect that if gravity is localized, then the corresponding Higgs mechanism
ensures absence of propagating negative-norm states.}.

\subsection{Comments}

{}Before we end our discussion, we 
would like to make a few comments. First, the aforementioned 
difficulties with the higher derivative terms might be under better control
if we consider a supersymmetric setup. However, attempts to construct 
non-singular smooth supersymmetric domain walls in the context of, say, 
$D=5$ ${\cal N}=2$ gauged supergravity have not been
successful so far (see, {\em e.g.}, \cite{BC,KL,BG}\footnote{In a recent
paper \cite{Behrndt} a construction of such a domain wall was reported. 
However, as was pointed out in \cite{Behrndt}, the corresponding solution
only exists if one {\em ad hoc} freezes one of the scalar fields whose 
dynamics is such that the domain wall is actually singular.}). 
In particular, typically one searches for solutions interpolating between
two local AdS minima of the scalar potential. However, as was argued in
\cite{ZW,Gib}, on general grounds potentials with more then one AdS minima 
are not expected to exist in $D=5$ ${\cal N}=2$ gauged 
supergravity\footnote{Note, however, that such domain walls 
have been constructed within the framework of $D=4$ ${\cal N}=1$ 
supergravity - see \cite{cvet} for a review.}. 

{}A possible way around this difficulty could be to consider potentials
with, say, no local minima at all. As we discussed in the previous sections,
such potentials admit non-singular solutions subject to certain conditions.
Understanding whether such potentials can arise in $D=5$ ${\cal N}=2$ gauged 
supergravity (that is, if they avoid the general ``no-go'' arguments of
\cite{ZW,Gib}) is beyond the scope of this paper. At any rate, it would be
interesting if one could construct a supersymmetric infinite tension domain
wall, or else prove that such solutions do not exist in $D=5$ ${\cal N}=2$ 
gauged supergravity\footnote{For a recent analysis of general $D=5$ ${\cal 
N}=2$ gauged supergravity models, see \cite{ceresole}.}.

{}The second comment concerns the ``self-tuning'' approach of \cite{ASKS,KSS}
to the cosmological constant problem. In \cite{Nilles} it was argued that
fine-tuning in these solutions is hidden in singularities. In \cite{erlich}
it was argued that in such solutions one either has a naked singularity or
fine-tuning. Here we would like to point out that, as we have already 
mentioned, the singularities in these solutions are unphysical. In particular,
it does not seem consistent to truncate the space in the $y$ direction at
such singularities. At any rate, the self-tuning property would only imply that
the four-dimensional cosmological constant is not affected by the quantum 
effects on the ``brane''. However, once bulk contributions are included, one
seems to lose control over the cosmological constant. In fact, a setup where
pure brane contributions to the cosmological constant vanish at least in the
conformal/large $N$ limit has been known \cite{BKV}, but as usual there too
one loses control over the cosmological constant once the coupling to gravity 
is included.    

\acknowledgments

{}I would like to thank Gregory Gabadadze,
Martin Ro{\v c}ek, Gary Shiu, Stefan
Vandoren and Peter van Nieuwenhuizen 
for discussions, and especially Gia Dvali for valuable discussions
and collaboration at some stages of this work. 
This work was supported in part by the National Science Foundation.
I would also like to thank Albert and Ribena Yu for financial support.


\begin{references}

\bibitem{early} 
V. Rubakov and M. Shaposhnikov, Phys. Lett. {\bf B125} (1983) 136.

\bibitem{BK}
A. Barnaveli and O. Kancheli, Sov. J. Nucl. Phys. {\bf 52} (1990) 576.

\bibitem{polchi} J. Polchinski, Phys. Rev. Lett. {\bf 75} (1995) 4724.

\bibitem{witt} P. Ho{\u r}ava and E. Witten, Nucl. Phys. {\bf B460} (1996)
506; Nucl. Phys. {\bf B475} (1996) 94;\\
E. Witten, Nucl. Phys. {\bf B471} (1996) 135.

\bibitem{lyk} J. Lykken, Phys. Rev. {\bf D54} (1996) 3693.

\bibitem{shif} G. Dvali and M. Shifman, Nucl. Phys. {\bf B504} (1997) 127;
Phys. Lett. {\bf B396} (1997) 64.

\bibitem{TeV} N. Arkani-Hamed, S. Dimopoulos and G. Dvali, 
Phys. Lett. {\bf B429} (1998) 263; Phys. Rev. {\bf D59} (1999) 086004.

\bibitem{dienes} K.R. Dienes, E. Dudas and T. Gherghetta, Phys. Lett. 
{\bf B436} (1998) 55; Nucl. Phys. {\bf B537} (1999) 47; hep-ph/9807522;\\
Z. Kakushadze, Nucl. Phys. {\bf B548} (1999) 205; Nucl. Phys.
{\bf B552} (1999) 3;\\
Z. Kakushadze and T.R. Taylor, Nucl. Phys. {\bf B562} (1999) 78.

\bibitem{3gen} Z. Kakushadze, Phys. Lett. {\bf B434} (1998) 269; 
Nucl. Phys. {\bf B535} (1998) 311.

\bibitem{anto} I. Antoniadis, N. Arkani-Hamed, S. Dimopoulos and G. Dvali,
Phys. Lett. {\bf B436} (1998) 257.

\bibitem{ST} G. Shiu and S.-H.H. Tye, Phys. Rev. {\bf D58} (1998) 106007.

\bibitem{BW} Z. Kakushadze and S.-H.H. Tye, Nucl. Phys. {\bf B548} (1999) 180.

\bibitem{Visser} M. Visser, Phys. Lett. {\bf B159} (1985) 22;\\
P. van Nieuwenhuizen and N.P. Warner, Commun. Math. Phys. {\bf 99} (1985)
141.

\bibitem{RS} L. Randall and R. Sundrum, Phys. Rev. Lett. {\bf 83} (1999)
3370; Phys. Rev. Lett. {\bf 83} (1999) 4690.

\bibitem{Gog} M. Gogberashvili, hep-ph/9812296; Europhys. Lett. {\bf 49} 
(2000) 396.

\bibitem{many} N. Arkani-Hamed, S. Dimopoulos, G. Dvali and N. Kaloper,
Phys. Rev. Lett. {\bf 84} (2000) 586.

\bibitem{ver} H. Verlinde, hep-th/9906182.

\bibitem{giga} B. Bajc and G. Gabadadze, hep-th/9912232.

\bibitem{weinberg} S. Weinberg, Rev. Mod. Phys. {\bf 61} (1989) 1.

\bibitem{DS} G. Dvali and M. Shifman, Phys. Lett. {\bf B454} (1999) 277.

\bibitem{Freedman} O. DeWolfe, D.Z. Freedman, S.S. Gubser and A. Karch,
hep-th/9909134.

\bibitem{Skenderis} K. Skenderis and P.K. Townsend, 
Phys. Lett. {\bf B468} (1999) 46.

\bibitem{gubser} S.S. Gubser, hep-th/0002160.

\bibitem{ASKS} N. Arkani-Hamed, S. Dimopoulos, N. Kaloper and R. Sundrum, 
hep-th/0001197.

\bibitem{KSS} S. Kachru, M. Schulz and E. Silverstein, hep-th/0001206.

\bibitem{erlich} C. Csaki, J. Erlich, C. Grojean and T. Hollowood,
hep-th/0004133.

\bibitem{KSS1} S. Kachru, M. Schulz and E. Silverstein, hep-th/0002121.

\bibitem{nihei} T. Nihei, Phys. Lett. {\bf B465} (1999) 81.

\bibitem{kaloper} N. Kaloper, Phys. Rev. {\bf D60} (1999) 123506.

\bibitem{odintsov} S. Nojiri, O. Obregon and S.D. Odintsov, hep-th/0005127.

\bibitem{GT} J. Garriga and T. Tanaka, hep-th/9911055.

\bibitem{giddings} S.B. Giddings, E. Katz and L. Randall, JHEP {\bf 0003}
(2000) 023.

\bibitem{DGP} G. Dvali, G. Gabadadze and M. Porrati, hep-th/0002190.

\bibitem{EW} E. Witten, hep-ph/0002297.

\bibitem{DVALI} G. Dvali, hep-th/0004057.

\bibitem{GRS} R. Gregory, V.A. Rubakov, S.M. Sibiryakov, hep-th/0002072;
hep-th/0003045.

\bibitem{CEH} C. Csaki, J. Erlich and T.J. Hollowood, hep-th/0002161;
hep-th/0003020.

\bibitem{DGP1} G. Dvali, G. Gabadadze and M. Porrati, hep-th/0003054;
hep-th/0005016.

\bibitem{Zwiebach} B. Zwiebach, Phys. Lett. {\bf B156} (1985) 315.

\bibitem{Zumino} B. Zumino, Phys. Rept. {\bf 137} (1986) 109.

\bibitem{Zee} I. Low and A. Zee, hep-th/0004124.

\bibitem{olindo} O. Corradini and Z. Kakushadze, to appear.

\bibitem{BC} K. Behrndt and M. Cveti{\v c}, hep-th/9909058; hep-th/0001159.

\bibitem{KL} R. Kallosh and A. Linde, hep-th/0001071.

\bibitem{BG} K. Behrndt and S. Gukov, hep-th/0001082. 

\bibitem{Behrndt} K. Behrndt, hep-th/0005185.

\bibitem{ZW} M. Wijnholt and S. Zhukov, hep-th/9912002.

\bibitem{Gib} G.W. Gibbons and N.D. Lambert, hep-th/0003197.

\bibitem{cvet} M. Cveti{\v c} and H. Soleng, Phys. Rept. {\bf 282}
(1997) 159.

\bibitem{ceresole} A. Ceresole and G. Dall'Agata, hep-th/0004111.

\bibitem{Nilles} S. Forste, Z. Lalak, S. Lavignac and H.P. Nilles, 
hep-th/0003152.

\bibitem{BKV} S. Kachru and E. Silverstein, Phys. Rev. Lett. {\bf 80}
(1998) 4855;\\
A. Lawrence, N. Nekrasov and C. Vafa, Nucl. Phys. {\bf B533} (1998) 199;\\
M. Bershadsky, Z. Kakushadze and C. Vafa, Nucl. Phys. {\bf B523} (1998) 59;\\
Z. Kakushadze, Nucl. Phys. {\bf B529} (1998) 157;\\
P.H. Frampton and C. Vafa, hep-th/9903226. 

\end{references}
\end{document}